\documentclass[preprint,showpacs,preprintnumbers,amsmath,amssymb]{revtex4}

\usepackage{graphicx}
\usepackage{dcolumn}
\usepackage{bm}
\usepackage{graphics}
\usepackage{color}
\usepackage{color}
\usepackage{subfigure}
\usepackage{epsfig}

\begin{document}
\preprint{}

{\Large\bf
\begin{center}
%
Final-State Interactions in the Superscaling Analysis of
Neutral-Current Quasielastic Neutrino Scattering
%
\end{center}
}
\vspace{1\fill}
\begin{center}
{\large
M.C. Mart\'{\i}nez$^{1}$, J.A. Caballero$^{2}$, T.W. Donnelly$^{3}$,
and J.M. Ud\'{\i}as$^{1}$
}
\end{center}

\date{\today}

\begin{small}
\begin{center}
$^{1}${\sl
Grupo de F\'{\i}sica Nuclear,
Departamento de F\'\i sica At\'omica, Molecular y Nuclear \\
Universidad Complutense de Madrid, E-28040  Madrid, SPAIN
}\\[2mm]
$^{2}${\sl
Departamento de F\'\i sica At\'omica, Molecular y Nuclear \\
Universidad de Sevilla, Apdo. 1065, E-41080 Sevilla, SPAIN
}\\[2mm]
$^{3}${\sl
Center for Theoretical Physics, Laboratory for Nuclear Science
and Department of Physics\\
Massachusetts Institute of Technology,
Cambridge, MA 02139, USA
}\\[2mm]

\end{center}
\end{small}

\kern 1. cm \hrule \kern 3mm

\begin{small}
\noindent

Effects of strong final-state interactions in the superscaling
properties of neutral-current quasielastic neutrino cross sections
are investigated using the Relativistic Impulse Approximation as
guidance. First- and second-kind scaling are analyzed for neutrino
beam energies ranging from 1 to 2 GeV for the cases of $^{12}$C,
$^{16}$O and $^{40}$Ca. Different detection angles of the outgoing
nucleon are considered in order to sample various nucleon energy
regimes. Scaling of the second kind is shown to be very robust.
Validity of first-kind scaling is found to be linked to the
kinematics of the process. Superscaling still prevails even in the
presence of very strong final-state interactions, provided that some
kinematical restrains are kept, and the conditions under which
superscaling can be applied to predict neutral-current quasielastic
neutrino scattering are determined.



\vspace{2mm}

\kern 1mm

\end{small}

\kern 2mm \hrule \kern 1cm

\maketitle

\section{Introduction}

 Accurate predictions for neutrino-nucleus cross sections are
needed in the analyses of on-going and future experimental studies
of neutrino reactions and neutrino oscillations~\cite{neut_exp} at
intermediate energies. One option is to rely on direct modeling of
the neutrino-nucleus interaction. To date this is the choice for
essentially all neutrino event generators employed in accelerator
and astroparticle neutrino experiments, where the relativistic Fermi
gas (RFG) is usually incorporated as a standard tool. Proceeding in
this way, one should keep in mind that any model for
neutrino-induced reactions should first succeed in comparisons with
the available high-quality inclusive electron cross section data,
since typically the various semi-leptonic cross sections are closely
related. A large variety of models, including the RFG, are not
successful in reproducing electron scattering data when agreement to
better than 20--30\% is desired. In order to avoid the nuclear
uncertainties inherent in any neutrino-nucleus reaction description,
the authors in \cite{Amaro:2004bs} have proposed the idea of
profiting from the extensive knowledge on nuclear dynamics acquired
from electron scattering experiments in order to predict inclusive
charged-current neutrino-nucleus cross sections. The connection
between the two electroweak processes is done by means of the
superscaling analysis.  The successful application of the
phenomenological SuperScaling Approach
(SuSA)~\cite{Amaro:2004bs,Caballero:2005sj,Amaro:2005dn,JA06,Amaro:2006if,Antonov:2006md,Martini:2007jw,Amaro:2006tf,Caballero:2007tz,Amaro:2006pr,Antonov2007_1}
has motivated us to revisit the concept of scaling and study its
validity when applied to neutral-current neutrino cross
sections.

It is very well established that the large amount of inclusive
electron scattering data manifests scaling behaviour at high
momentum transfer for excitation energies falling below the
quasielastic (QE)
peak~\cite{Day:1990mf,Donnelly:1998xg,Donnelly:1999sw,Maieron:2001it,Barbaro:1998gu}.
These $(e,e')$ data (particularly those coming from the analysis
of the longitudinal response), when appropriately organized, show
reasonably good scaling of first kind (no dependence on the
momentum transfer $q$) and excellent scaling of second kind (no
dependence on the particular nuclear species). They are said to
superscale. As an outcome of this behaviour, a phenomenological
scaling function has been directly derived from the longitudinal
data~\cite{Donnelly:1998xg,Maieron:2001it}. This scaling function
contains the relevant information about the initial- and
final-state nuclear dynamics explored by the probe, in this case,
an electron. Recently, the scaling analysis has also been extended
to inelastic responses in the region of the delta peak and even
beyond~\cite{Amaro:2004bs,Barbaro:2003ie}, and scaling ideas have
also been applied to processes involving hadronic probes~\cite{peterson}. It is
important to notice that superscaling is a general phenomenon
exhibited by Nature (although not perfectly). Essentially it shows
up upon dividing the electron-nucleus cross section data at
sufficiently high $q$ by the corresponding electron-nucleon cross
section, and plotting the results against a properly chosen
scaling variable, which can be derived using simple kinematical
considerations. Superscaling of data is thus independent of any
modeling of the reaction, although it is a characteristic of the
inclusive electron scattering that any ``reliable'' model should
be able to reproduce. For instance, it is well known that,
although the relativistic Fermi gas exhibits perfect
superscaling~\cite{Alberico:1988bv} (it has even inspired a
particular scaling variable), the scaling function it predicts
differs from the experimental result. As a matter of fact, only a
few models to date can accurately reproduce the experimental
$(e,e')$ scaling function. One of these is based on the
Relativistic Impulse Approximation when strong relativistic
mean-field potentials are used to describe the bound and
ejected nucleon wave functions. This point will be discussed at
length later.

The kinematics involved in $(e,e')$ and $(\nu,\mu)$ reactions are
rather similar, {\it i.e.,} in both cases the scattered lepton is
detected, and both the energy and momentum transferred by the probe
to the nucleus are thus known. Accordingly, it is plausible to
expect that the two probes explore the nucleus in a similar way, and
consequently both electron and charged-current neutrino scattering share
 the same
universal scaling function. This is the main approximation adopted
by SuSA, where the scaling function is determined using {\it
longitudinal} electron scattering data and then carried forward to
make predictions for neutrino-induced
processes~\cite{Amaro:2004bs,Amaro:2006tf}. These phenomenological
predictions, which incorporate nuclear information provided directly
by the analysis of $(e,e')$ experimental data, are believed to be
more robust and ``reliable'' than those coming from direct modeling.

A question arises naturally, namely, can these superscaling ideas be
also applied to neutral-current neutrino-nucleus cross sections? Can we get
reliable predictions for neutral-current processes based on the
 phenomenological
electron superscaling function? These neutrino processes play an
important role in the determination of the strange quark
contribution to the nucleon spin. Furthermore, neutral-current
 reactions are also
relevant for oscillation experiments --- for instance, it is
expected that they contribute as the third most important event type
for the MiniBooNE experiment at Fermilab~\cite{neut_exp}.  As in the
case of charged-current processes, neutral-current neutrino-nucleus 
cross section predictions
based on scaling ideas, when possible, are clearly demanded.
However, the extension of the scaling analysis and the application
of SuSA ideas to neutral-current reactions is not as straightforward as in the
case of charged-current processes.
 Focusing on the quasielastic region, in the case of neutral-current
 (NC) processes the scattered neutrino is not detected and identification
of the NC event may be made by observing a nucleon ejected from the
nucleus without finding evidence of final charged leptons. Even
measuring the nucleon energy and momentum, the transferred energy
and momentum at the leptonic vertex will remain unknown. Hence, the
kinematics for NC are clearly different from those corresponding to
electron and charged-current neutrino scattering processes. This makes the
applicability of scaling arguments to neutral-current reactions less
obvious~\cite{Amaro:2006pr}. Given the basic differences in the two
types of kinematics, it is first necessary to establish the validity
or not of scaling ideas for neutral-current neutrino scattering. Secondly, the
phase-space regions explored in the two processes (neutral-current versus
electron and charged-current neutrino scattering) are different, and they could
display distinct sensitivities to the many-body physics underlying
the scaling function. Thus, the idea of using the experimental
electron scaling function to predict neutral-current neutrino cross sections,
 as
was done for charged-current processes, requires a more in-depth analysis. In
other words, prior to any extension of SuSA analyses to predict neutral-current
 neutrino scattering in the quasielastic region, one needs to be sure if
superscaling also holds for NC neutrino-nucleus scattering, and
moreover, if the phenomenological $(e,e')$ scaling function can
safely be used to predict NC cross sections, in spite of the
intrinsic differences between the two processes.

In order to proceed, the ideal strategy would be to count on
sufficient neutral-current neutrino-nucleus experimental data to perform a
similar scaling analysis to the one used for the electron case, and
compare the NC scaling function (if any) with the electron one.
Unfortunately, neutral-current data are too scarce to allow for a meaningful
scaling analysis, and one has to rely on neutrino-nucleus models
that are expected to mimic the behaviour that data should show.

Keeping this in mind, the natural starting point for examining
scaling properties of neutral-current neutrino scattering is the relativistic
 Fermi gas. This model
is known to exhibit perfect superscaling when applied to inclusive
quasielastic electron and charged-current neutrino scattering. Such a study was recently
undertaken for neutral-current neutrino reactions in~\cite{Amaro:2006pr}, where
it was shown how to extend the scaling ideas to NC processes (within
the RFG model), illustrating the results for scattering of 1 GeV
neutrinos from $^{12}$C. Additionally, the universality property of
the scaling function was assumed in \cite{Amaro:2006pr}, and hence
NC neutrino cross sections were predicted by making use of the
averaged phenomenological scaling function extracted from the
analysis of $(e,e')$ data. The same idea of universality of the
scaling function was exploited in~\cite{Antonov2007_1}, where the
coherent density fluctuation model scaling function was used
to predict neutral-current neutrino cross sections, and results were compared
with those based on the phenomenological $(e,e')$ scaling function.
A first obstacle to assessing the likelihood of being able to
predict NC cross sections in this way is that the RFG, although
providing a good framework to start with, is surely too simple to
describe satisfactorily the behavior showed by NC data if
uncertainties of less than 20--30$\%$ are desired. Although the RFG
$(e,e')$ response by definition exhibits perfect
superscaling~\cite{Alberico:1988bv}, it lacks important dynamical
effects, hence providing responses which are not fully in accord
with the magnitude and shape of the experimental scaling function.
Indeed, it has been shown that strong final-state interactions (FSI)
are needed to describe successfully the magnitude and shape of the
superscaled data, introducing also small deviations from perfect
superscaling behaviour. It is expected that FSI also modify the neutral-current
 neutrino-nucleus cross sections in a significant way, and thus it is
necessary to infer how FSI may affect the scaling properties.

In this work, which is complementary to that in~\cite{cris07}, we
address this issue: we perform a systematic analysis of the effects
of strong final-state interactions in the superscaling properties of quasielastic neutral-current neutrino-nucleus scattering within the context of the Relativistic Impulse
 Approximation (RIA), based on strong relativistic mean-field (RMF) potentials for both the bound and the ejected nucleons.
Several arguments motivate our choice of this particular model to
illustrate the possible effects of FSI. The RIA model has been
extensively and successfully applied in investigations of exclusive
electron scattering reactions~\cite{Udias}. Furthermore, the RIA
approach has been shown to superscale when applied to quasielastic inclusive
electron and charged-current neutrino scattering, giving rise to a ``unique''
scaling function, with relatively mild scaling violations. This result
reinforces the idea of the existence of a universal scaling function
that is valid for both electron and charged-current neutrino
probes~\cite{Caballero:2005sj,JA06}. Finally, the most important
reason for the choice of the RIA, as well as its corresponding
semi-relativistic version~\cite{Amaro:2006if}, is that it is capable
of reproducing the shape and magnitude of the experimental scaling
curve extracted from QE $(e,e')$ data, something that has proven to
be elusive for other theoretical models. In particular, the
asymmetric shape exhibited by the experimental scaling function with
a significant tail extending to high values of the transfer energy,
is accurately reproduced by the RIA, provided that FSI are described
with the relativistic mean field approach~\cite{Caballero:2005sj,JA06}.
 For all these
reasons, and, being aware of the scarcity of neutral-current neutrino data, we
are led to accept the validity of the RIA with final-state interactions
 treated through
the use of the RMF as a realistic model for the description of
inclusive scattering and believe that it can serve as a reliable
tool to illustrate possible FSI effects on the scaling properties of
quasielastic neutral-current neutrino cross sections.

The paper is organized as follows. In Section II we present a
brief summary of the formalism required to treat NC
neutrino-nucleus scattering in the RIA. A general discussion of
the superscaling phenomenon is also provided, showing the basic
expressions to be used in the analysis. In Section III we discuss
the results. First, we focus on differential cross sections and
response functions for neutrino- and antineutrino-induced proton
and neutron knockout. Several kinematical situations of interest
are examined. Second, we evaluate the scaling function and perform
a separate study of the scaling properties --- scaling of the
first and second kinds --- paying special attention to FSI
effects. Results are illustrated for various choices of kinematics
and for three nuclei, $^{12}$C, $^{16}$O and $^{40}$Ca. A
comparison between the RIA-RMF NC and the phenomenological
electron scaling functions is also performed. Finally, in Section
IV we summarize our main conclusions.

\section{Formalism for NC neutrino-nucleus scattering}
\label{sec:formalism}

Following the general procedure for superscaling analyses, we start
by evaluating inclusive QE neutral-current neutrino cross sections within
the RIA model. We assume the inclusive cross section to be
given as the integrated semi-inclusive one-nucleon (proton or
neutron) knockout cross sections. This approximation, which is
implicit in scaling analyses, has been shown to work successfully in
the kinematic region dominated by quasielastic scattering. In other words, we
construct the inclusive $A(\nu,N)\nu'A-1$ cross section, within the RIA model,
by integrating
the $A(\nu,\nu'N)A-1$ cross section over the unobserved scattered
neutrino variables.

\subsection{Relativistic Impulse Approximation (RIA)}
\label{sec:riaformalism}

The RIA model has been used to describe neutral-current neutrino-nucleus
reactions in previous work~\cite{Alb97,Martinez:2005xe}. Here we
simply summarize those aspects which are of most relevance for
later discussion of the scaling properties.

\begin{figure}
{\par\centering
\resizebox*{0.9\textwidth}{0.4\textheight}{\rotatebox{0}
{\includegraphics{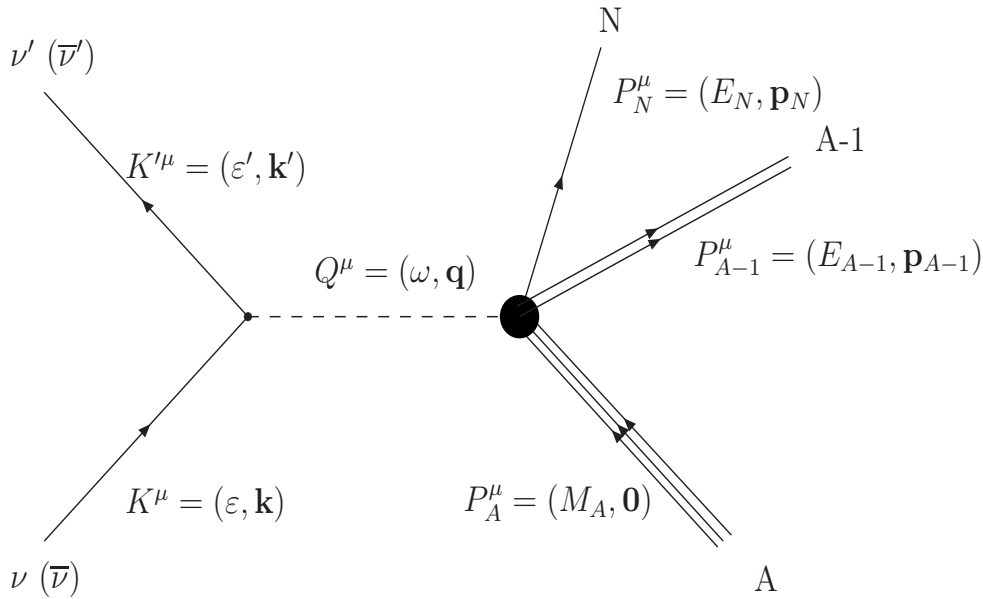}}} \par} \caption{Feynman diagram for quasielastic NC neutrino-nucleus scattering in the first Born approximation}
\label{fig:kinema} 
\end{figure}

The first basic assumption of the RIA is that the process occurs
through the exchange of a single vector boson; this is known as
the first Born Approximation. In this approach, the leading-order
exclusive quasielastic cross section is generated by the Feynman amplitude
associated with the diagram shown in Fig.~\ref{fig:kinema}. Here,
a neutrino scatters off an A-body nucleus via the exchange of a
$Z^0$. In the scattering process, a nucleon is knocked out,
leaving behind an (A-1)-body daughter nucleus, generally in an
excited state. The kinematical variables can be inferred from the
figure.

The RIA also assumes the impulse approximation, {\it i.e.,}
the incident neutrino interacts with only one nucleon which is
subsequently emitted. The nuclear current is written as a sum of
single-nucleon currents. Then, the transition matrix elements from
which the cross section is computed can be cast in the following
form:
\begin{equation}
\label{eq:relcurrent}
  \langle J^{\mu}
  \rangle = \int
  d\mathbf{r} \; \overline{\phi}_F(\mathbf{r})\hat{J}^{\mu}(\mathbf{r})
  e^{i\mathbf{q}\cdot\mathbf{r}}\phi_{B}(\mathbf{r}) \; ,
 \end{equation}
where $\phi_{B}$ and $\phi_F$ are relativistic bound-state and
scattering wave functions, respectively, and $\hat{J}^{\mu}$ is
the relativistic one-body current operator modeling the coupling
between the virtual $Z^0$ and a bound nucleon
(see~\cite{Alb97,Martinez:2005xe} for details concerning the
operator and nucleon form factors; in all results presented in the
next section we have not allowed for strangeness content in the
nucleon). We describe the bound nucleon states as self-consistent
Dirac-Hartree solutions, derived within a relativistic mean-field
approach using a Lagrangian containing $\sigma$, $\omega$ and
$\rho$ mesons~\cite{boundwf}.

Ignoring all distortions due to final-state interactions
 leads to the description of
the scattering wave function for the outgoing nucleon as a
relativistic plane wave. This is known as the Relativistic
Plane-Wave Impulse Approximation (RPWIA)~\cite{Caballero:1997gc,
Caballero:1998ip}, which obviously entails a simplified
description of the reaction mechanism. On the contrary, in the
present work, when accounting for final-state interactions between
the ejected nucleon and the residual nucleus, the outgoing nucleon
wave function is computed using the same relativistic mean field
employed to describe the initial bound states. We denote this
approach as
RMF~\cite{Alb97,Chiara03,Martinez:2005xe,JA06,Caballero:2005sj,Jin92}.

Using these ingredients, we evaluate the six-differential cross
section $d^6\sigma/d\varepsilon'd\Omega_{k'}dE_Nd\Omega_N$. The
inclusive cross section is obtained by integrating over the three
momenta of the undetected particles. In neutral-current neutrino scattering the
outgoing nucleon is assumed to be the only particle detected in
the final state; hence one integrates over the scattered neutrino
variables $\varepsilon'$ and $\Omega_{k'}$. A sum over all
single-particle states from which the nucleon may originate is
also performed.

As can be inferred from the way we are proceeding, processes
containing other particles in addition to the nucleon in the final
state, including multinucleon knockout and pion production, are not
explicitly taken into account. Within the context of the RMF
(mean-field-based model) only one-body processes are explicitly
considered and the inclusive strength is exclusively built out of
impulse approximation single-nucleon knockouts. Note, however, that
the one-body contribution from multinucleon knockout is fully
incorporated into the RMF model, via the self-energy of the
propagating nucleon. In other words, the RMF approach includes all
kind of rescattering (elastic and inelastic) with the remaining
nucleons. Thus, in this mean-field picture the redistribution of the
strength and multi-nucleon knockout is completely attributed to final-state
 interactions 
and not to explicit correlations, which are in fact incorporated
within the relativistic mean field in an effective way. One must notice that the
relativistic mean field, due to the presence of strong scalar and
vector potentials combining repulsive and attractive interactions
that provide the correct saturation properties for nuclear matter,
has more flexibility to incorporate correlations in an effective way
than does the non-relativistic mean-field approach~\cite{plohl}.

The adequacy of the potential to be
employed for inclusive scattering has been investigated in several works~\cite{hori,
chinn,meucci03,meucci04}. Besides the relativistic mean field,
approaches where one uses only the real part of the optical potential~\cite{Chiara03} or
where the full optical potential is employed but a formal summation based
on Green function formalism over all possible channels that contribute to
the inclusive scattering is done~\cite{hori,meucci03,meucci04}, have been used to describe
lepton-nucleus scattering under inclusive conditions. The comparison of
results from these approaches show that they differ by a few percent
and in general they compare equally well to existing electron scattering 
data. Thus, in this work we present only results obtained within the relativistic mean field.

\subsection{Scaling and superscaling analysis}
\label{sec:scalingform}

In the case of electron and charged-current neutrino scattering, the general
procedure to get the scaling function is to divide the inclusive
differential cross section by the single-nucleon $eN$ or $\nu N$
cross section, weighted by the number of protons and neutrons
involved in the process. If the function obtained in this way,
when plotted against a properly chosen variable (the scaling
variable), is seen to depend weakly on the momentum transfer, one
observes scaling of first kind. Additionally, both the scaling
function and the scaling variable can be made dimensionless via a
characteristic momentum scale for the chosen nucleus. If the
scaling function so obtained does not depend on the particular
nuclear species involved, one talks about scaling of the second
kind. When both types of scaling occur, the data or the
calculations are said to superscale. One should notice the
important role played by the kinematics in the general derivation
of scaling for both electron and charged-current neutrino scattering. Two
factors seem to be essential for scaling to occur: i) the
transferred four-momentum $Q^\mu$ is fixed in the process, and ii)
the missing-energy missing-momentum region accessible to the
reaction is relatively confined. This is the situation for the
$t$-channel processes (electron and charged-current neutrino scattering).

Being aware of the essential differences mentioned in the
introduction between the kinematics involved in inclusive electron
or charged-current neutrino reactions and those for neutral-current
 neutrino processes, we
start the present study in the Relativistic Impulse Approximation
 reviewing briefly the general
discussion on the QE neutral-current kinematics and scaling properties
presented in \cite{Amaro:2006pr}. Summarizing, for neutral-current scattering
we assume the energy $\varepsilon$ of the incident neutrino to be
specified (in real situations there are usually no monochromatic
beams and an integral over the allowed energies folded with the
neutrino flux must be performed), and also assume the outgoing
nucleon energy $E_N$ to be known. Finally, the angle
$\theta_{kp_N}$ between the incident neutrino and the ejected
nucleon momentum is also given. Notice that, in contrast to
electron and charged-current neutrino reactions, the scattered lepton four
momentum $Q^\mu=K^\mu-K'^\mu$ is not fixed, making the analysis of
the scaling behaviour more difficult. As already presented
in~\cite{Amaro:2006pr}, one can introduce a corresponding constant
Lorentz invariant $Q'^2=(K^{\mu}-P_N^{\mu})^2=\omega'^2-q'^2$,
where $\omega'=\varepsilon-E_N$ and
$q'=|\mathbf{q}'|=|\mathbf{k}-\mathbf{p}_N|=\sqrt{\varepsilon^2+p_N^2-2\varepsilon
p_Ncos\theta_{kp_N}}$.

In~\cite{Amaro:2006pr} the influence of a non-constant $Q^\mu$ in
the derivation of scaling in the neutral-current case was thoroughly
investigated, concluding within the general framework of the relativistic
 Fermi gas 
model that scaling ideas still work properly for neutral-current
neutrino-nucleus processes. In this work we shall not repeat that
study, but rather we take it as our starting point. Our main aim
here is to go a step further by analyzing whether or not the
presence of strong final-state interactions
 affects the scaling properties of QE neutral-current
cross sections. Inspired by the RFG result, in this work we
construct the following function from the inclusive neutral-current neutrino
cross section evaluated in the relativistic impulse approximation,
\begin{eqnarray}
\label{eq:supers_f} f(q',\psi^u)\equiv k_A F(q',\psi^u)= k_A
\frac{\left[\frac{d\sigma}{d\Omega_Ndp_N}\right]}{\overline{\sigma}_{sn}^{NC}}\;
,
\end{eqnarray}
 where
$\overline{\sigma}_{sn}^{NC}$ is the effective neutral-current single-nucleon
cross section (given explicitly in Eq.~(20) and the appendix in
\cite{Amaro:2006pr}), and $\psi^u(q',\omega')$ is the
dimensionless scaling variable extracted from the RFG analysis to
be used for QE neutral-current kinematics (see Eqs.~(24)-(26) in~\cite{Amaro:2006pr}). The momentum $k_A$ introduced in
Eq.~(\ref{eq:supers_f}) states a characteristic momentum scale for
a given nuclear species, allowing one to define the dimensionless
scaling function $f(q',\psi^u)$. This scaling function can be
plotted versus $\psi^u$ for different neutrino beam energies (which
at fixed $\theta_{kp_N}$ and given $E_N$ means different $q'$ values) and
different nuclear species, thereby yielding a way of analyzing
first- and second-kind scaling.

It is important to note that the azimuthal angle $\phi_N$ between
the outgoing nucleon and the scattering plane (which contains both
the incident and the scattered neutrinos) will not cover the full
range from 0 to $2\pi$ (see discussion in~\cite{Amaro:2006pr}),
and hence there will be nonzero contributions in the inclusive neutral-
current
cross sections with detection of the outgoing nucleon (TL, TT, TL'
terms) which would vanish in the usual inclusive charged-current neutrino and
electron scattering reactions.

\section{Results}
\label{sec:results}

\subsection{Differential cross sections and responses}
\label{sec:crosssec}

\begin{figure}
{\par\centering
\resizebox*{1.1\textwidth}{0.55\textheight}{\rotatebox{270}
{\includegraphics{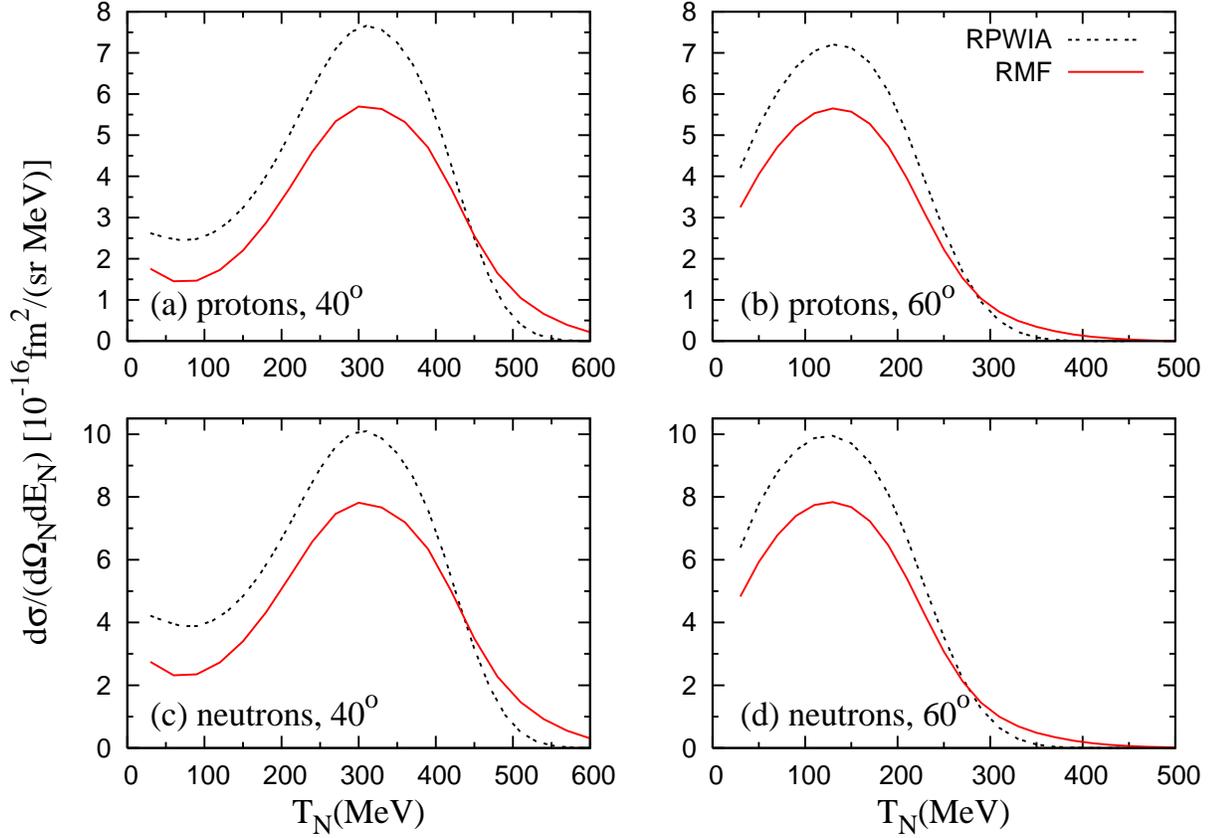}}} \par} \vspace{2cm} \caption{(Color online) QE
differential cross sections for NC neutrino scattering at 1 GeV from
$^{12}$C for proton (panels (a) and (b)) and neutron (panels (c) and (d))
knockout. Panels (a) and (c) correspond to $\theta_{kp_N}=40^o$
and panels (b) and (d) to 60$^o$. Results are given for the RPWIA
and the RMF-FSI description.} \label{fig:crossc12nu}
\end{figure}

We begin by evaluating final-state interaction effects on inclusive QE neutral-current neutrino
scattering. We first focus on the case of $^{12}$C, whose choice
is motivated not only by the relevance for present neutrino
oscillation experiments, but also because it facilitates the
comparison with the relativistic Fermi gas results presented in \cite{Amaro:2006pr}.
In Fig.~\ref{fig:crossc12nu} we present the differential cross
section $d\sigma/dE_Nd\Omega_N$ at incident neutrino energy
$\varepsilon=1$ GeV for $^{12}$C$(\nu,p)$ and $^{12}$C$(\nu,n)$ as
a function of the outgoing nucleon kinetic energy $T_N$. The two curves
in each panel correspond to different descriptions of the outgoing
nucleon wave function: RPWIA (dashed) and RMF potential (solid).
Results are shown for two representative values of
 the nucleon scattering angle, 40$^o$ (panels (a) and (c)) and 60$^o$ 
(panels (b) and (d)).

Final-state interactions lead to a decrease of the cross sections. Furthermore, RMF
results exhibit an increase of strength at large $T_N$. The
existence of this longer tail within the RMF approach is due to
the strong relativistic potentials used in describing the
final-state interaction. This is consistent with what was already
observed and discussed within the context of $(e,e')$ and charged-current
neutrino-nucleus cross sections, where a pronounced tail extending
to small final-state lepton kinetic energies also emerges from the
use of the relativistic mean-field potential~\cite{Caballero:2005sj,JA06}. The
magnitude of the scalar and vector potentials in RMF, which are
energy-independent, can shift significant strength to higher
nucleon kinetic energies~\cite{Caballero:2005sj,JA06}. This effect
of the strong RMF potential is the one that provides the
theoretical $(e,e')$ and $(\nu,\mu)$ RIA superscaling functions
with the correct asymmetry, namely, the one observed in the
experimental data at high energy transfers (positive values of the
 scaling variable).

\begin{figure}
{\par\centering
\resizebox*{1.1\textwidth}{0.55\textheight}{\rotatebox{270}
{\includegraphics{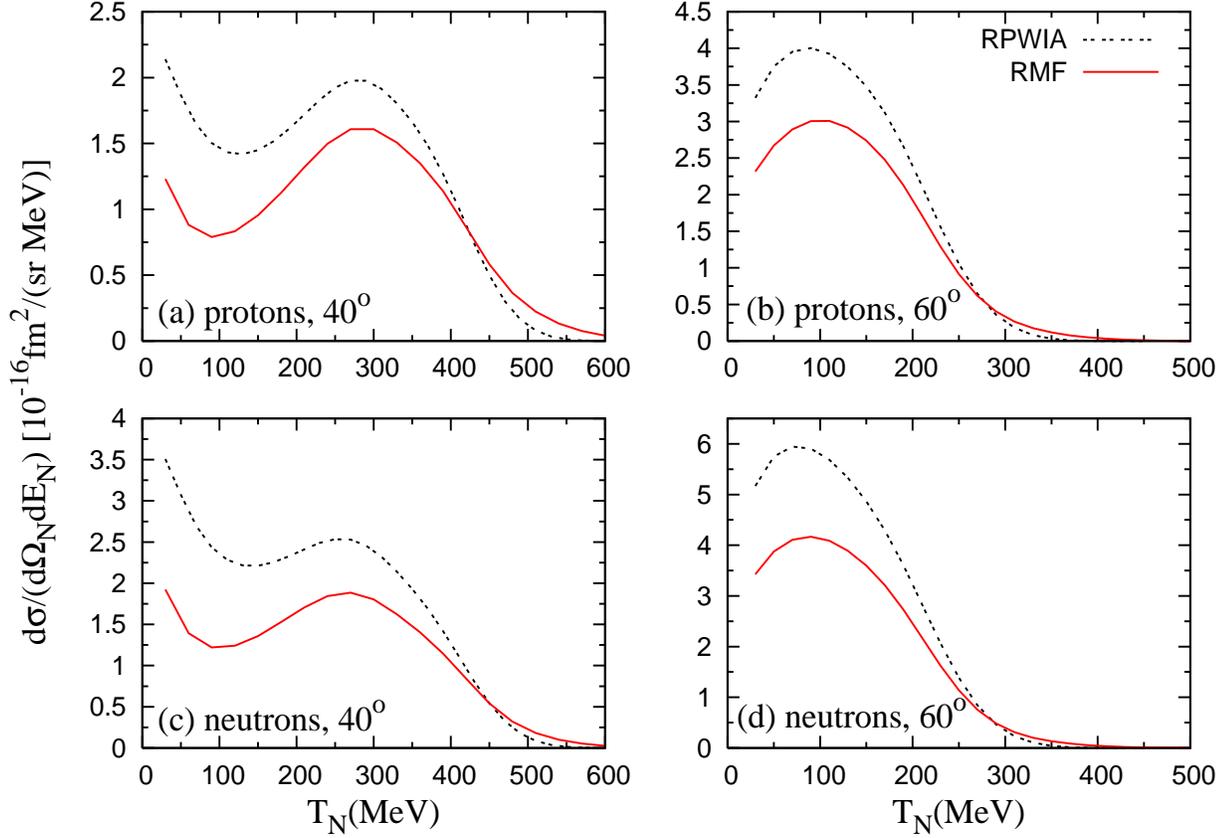}}} \par} \vspace{2cm}
\caption{(Color online) As in Fig.~\ref{fig:crossc12nu}, but now for
antineutrinos.} \label{fig:crossc12anu}
\end{figure}

As shown in Fig.~\ref{fig:crossc12anu}, final-state interactions have a similar influence
 on antineutrino proton and neutron knockout cross sections. Notice
that $\overline{\nu}$ cross sections are somewhat smaller,
particularly for the lower values of $\theta_{kp_N}$ (forward
scattering). Second, the long tail displayed for low $T_N$ and
observed at $\theta_{kp_N}=40^o$ is significantly more pronounced
in the antineutrino case. Note that the cross sections even change
their behaviour in this region; they increase significantly as
$T_N$ approaches zero.

In accordance with what was observed within the context of the RFG
in \cite{Amaro:2006pr}, a common feature of the two models, RPWIA
and RMF, is that the differential cross sections present a similar
shape for both proton and neutron emission, although with somewhat
different magnitude. Neutron ejection involves larger cross
sections, as one expects from the isovector dependence of the
coupling to the $Z^0$~\cite{Amaro:2006pr}.

\begin{figure}
{\par\centering
\resizebox*{1.1\textwidth}{0.55\textheight}{\rotatebox{270}
{\includegraphics{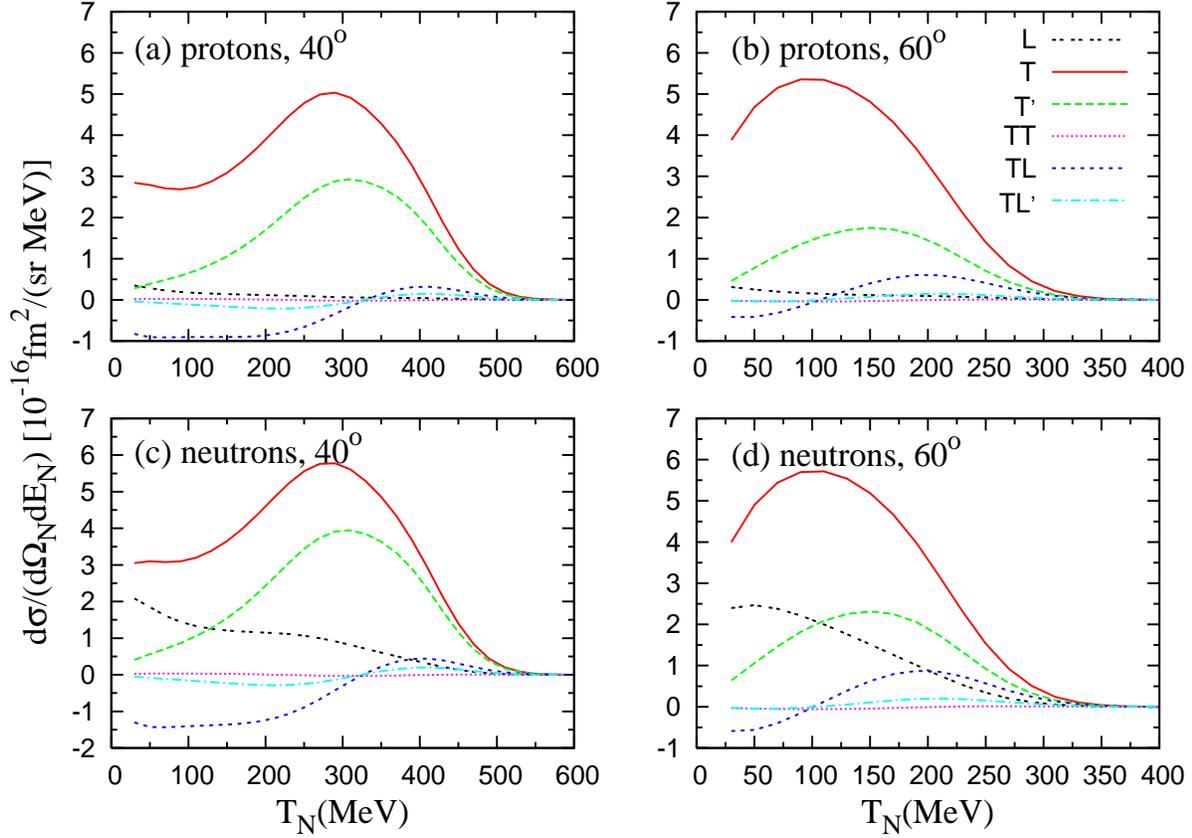}}} \par} \vspace{2cm}
\caption{(Color online) Response functions contributing to the RPWIA cross sections
in Fig.~\ref{fig:crossc12nu} for proton (panels (a) and (b)) and neutron
(panels (c) and (d)) knockout. Panels (a) and (c) correspond to
$\theta_{kp_N}=40^o$ and panels (b) and (d) to 60$^o$.}
\label{fig:responsesc12}
\end{figure}

The separate response functions contributing to the RPWIA cross
sections in Figs.~\ref{fig:crossc12nu} and ~\ref{fig:crossc12anu}
 are displayed in
Fig.~\ref{fig:responsesc12}. These RPWIA responses are very similar to those
presented for the relativistic Fermi gas model in \cite{Amaro:2006pr} at
$\theta_{kp_N}=60^o$ (see Fig.~6 in the earlier reference). This
result is indeed consistent with the minor role of nuclear
modeling expected at energies of the order of 1 GeV and
higher~\cite{Chiara03}.
Thus, most of the features concerning the responses at 60$^o$
signaled in \cite{Amaro:2006pr} apply also to our present
results within the relativistic impulse approximation: the purely transverse $T$ and $T'$ responses yield the
dominant contributions. In the case of neutron emission, the
purely longitudinal $L$ response produces also a significant
contribution, particularly for small values of the nucleon kinetic
energy. The $TL$ interference response also contributes to the
shape and magnitude of the cross sections. This applies not only
to the 60$^o$ results, but also to the 40$^o$ ones. Finally, the
$TT$ and $TL'$ response functions have an almost negligible effect
on the cross section for all cases.

Although not shown for brevity, the RMF responses have similar
trends to those for the RPWIA case, but with reduced magnitudes. In
particular, responses evaluated with the RMF potential vanish at
higher $T_N$, in accordance with the behaviour of the cross sections
shown in Fig.~\ref{fig:crossc12nu}.

\begin{figure}
{\par\centering
\resizebox*{0.9\textwidth}{0.9\textheight}{\rotatebox{270}
{\includegraphics{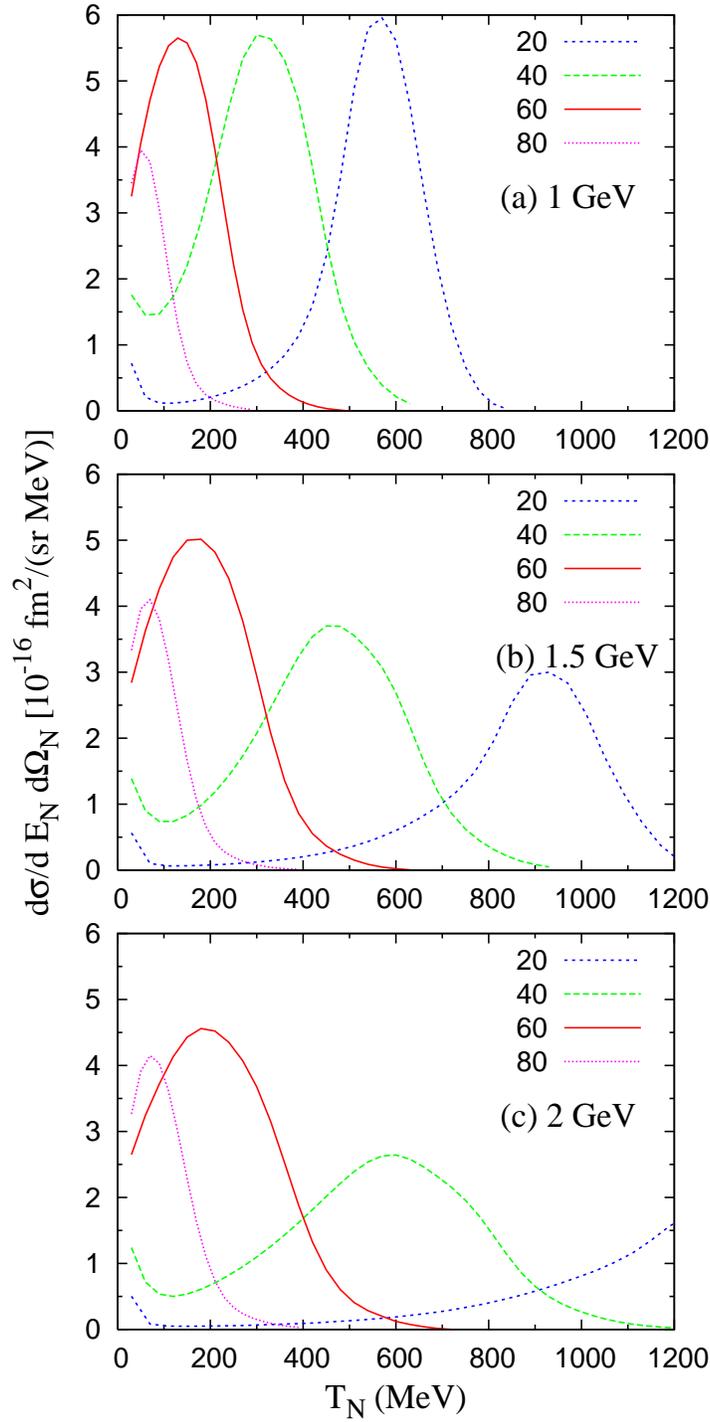}}} \par} \caption{(Color online) RMF differential
cross sections for the NC $^{12}$C$(\nu,p)$ reaction at various
beam energies and detection angles of the outgoing proton
$\theta_{kp_N}$. Panels (a), (b) and (c) correspond to
$\varepsilon =$ 1, 1.5 and 2 GeV, respectively. The different
curves in each panel are obtained by changing the value of
$\theta_{kp_N}$ from 20 to 80 degrees, as specified in the label.}
\label{fig:figura2}
\end{figure}

Let us now turn to the main goal of this section, namely the
behaviour of the QE neutral-current neutrino cross sections with respect to
changes in the kinematics. This is crucial for understanding the
outcomes of the scaling analysis discussed in next section. The
results in Fig.~\ref{fig:figura2} correspond to $^{12}$C$(\nu,p)$
differential cross sections within RMF for three neutrino beam
energies, 1, 1.5 and 2 GeV (panels (a), (b) and (c), respectively).
 For each energy, we study the variation of the
cross sections with the nucleon scattering angle. Several
observations can be made based on the figure:
\begin{itemize}
\item For a given neutrino energy, $\theta_{kp_N}$ determines the
value of the nucleon kinetic energy for which the cross section
presents its maximum. In general, small angles correspond to cross
sections peaked at higher values of kinetic energy. Hence,
changing the value of $\theta_{kp_N}$ allows us to explore
different regions of nucleon kinetic energy. \item For a given
$\theta_{kp_N}$, changing the neutrino beam energy means changing
the position and magnitude of the maximum of the cross section.
This change is significantly stronger at more forward angles.
\end{itemize}

\begin{figure}
{\par\centering
\resizebox*{0.6\textwidth}{0.27\textheight}{\rotatebox{270}
{\includegraphics{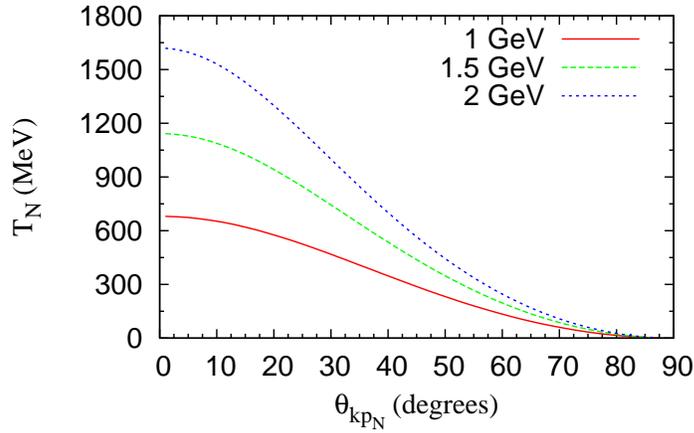}}} \par} \caption{(Color online) Relationship between 
the detection angle of the
outgoing proton and the value of the proton kinetic energy for the
NC $(\nu,p)$ reaction on free protons at rest. The
different curves show the results for three beam energies, 1, 1.5
and 2 GeV. } \label{fig:casolibre}
\end{figure}

The evolution of the QE neutral-current $^{12}$C$(\nu,p)$ cross section can be easily
understood from the consideration of the corresponding cross section from
free nucleons at rest. In such a case, energy and momentum conservation imposes
a relationship between the values of $\theta_{kp_N}$, the angle of the nucleon
with respect to the incident neutrino beam, and $T_N$, the energy of
the final nucleon. This relation is given in the form,
\begin{equation}
T_N=\frac{2 M_N \varepsilon^2 \cos^2{\theta_{kp_N}}}{(\varepsilon+M_N)^2-\varepsilon^2 \cos^2{\theta_{kp_N}}}.
\label{eq:tnfree}
\end{equation}
Results from Eq.~(\ref{eq:tnfree}) for several
values of the incoming neutrino energy, $\varepsilon$, are shown in
Fig.~\ref{fig:casolibre}. For bound nucleons, apart from the effect of
Fermi motion inside the nuclei that broadens the peak and
shifts it a bit with regard to the free-nucleon case, we expect
the maxima of the cross sections to be located approximately where
Eq.~(\ref{eq:tnfree}) predicts (see Fig.~\ref{fig:casolibre}). This explains
in a simple way the observations about the QE neutral-current $^{12}$C$(\nu,p)$ cross section specified above.
These results are thus directly linked to the kinematics of the process, and not to the
use of the RIA-RMF model. In particular,
the fact that the range of $T_N$ spanned at
fixed $\theta_{kp_N}$ for varying beam energy is reduced for large
angles as seen in Figs.~\ref{fig:figura2} and \ref{fig:casolibre} will have important
implications in the first-kind scaling properties of the cross sections.

\subsection{Scaling analysis}

In this section we present a general study of the scaling
properties of the QE neutral-current neutrino cross sections with strong final-state interactions.
 Our aim is twofold. First, we investigate
whether FSI effects may prevent (or not) the appearance of
superscaling. Second, if superscaling holds even in this case, we
study the conditions under which the SuSA approach (based on the
experimental superscaling $(e,e')$ response) can be used to
predict neutral-current neutrino-nucleus cross sections. As stated in the
introduction, we use the RIA-RMF model to illustrate the results,
although our findings are not intrinsically linked to the use of
this model, as will be clarified later.

\begin{figure}
{\par\centering
\resizebox*{1.1\textwidth}{0.55\textheight}{\rotatebox{270}
{\includegraphics{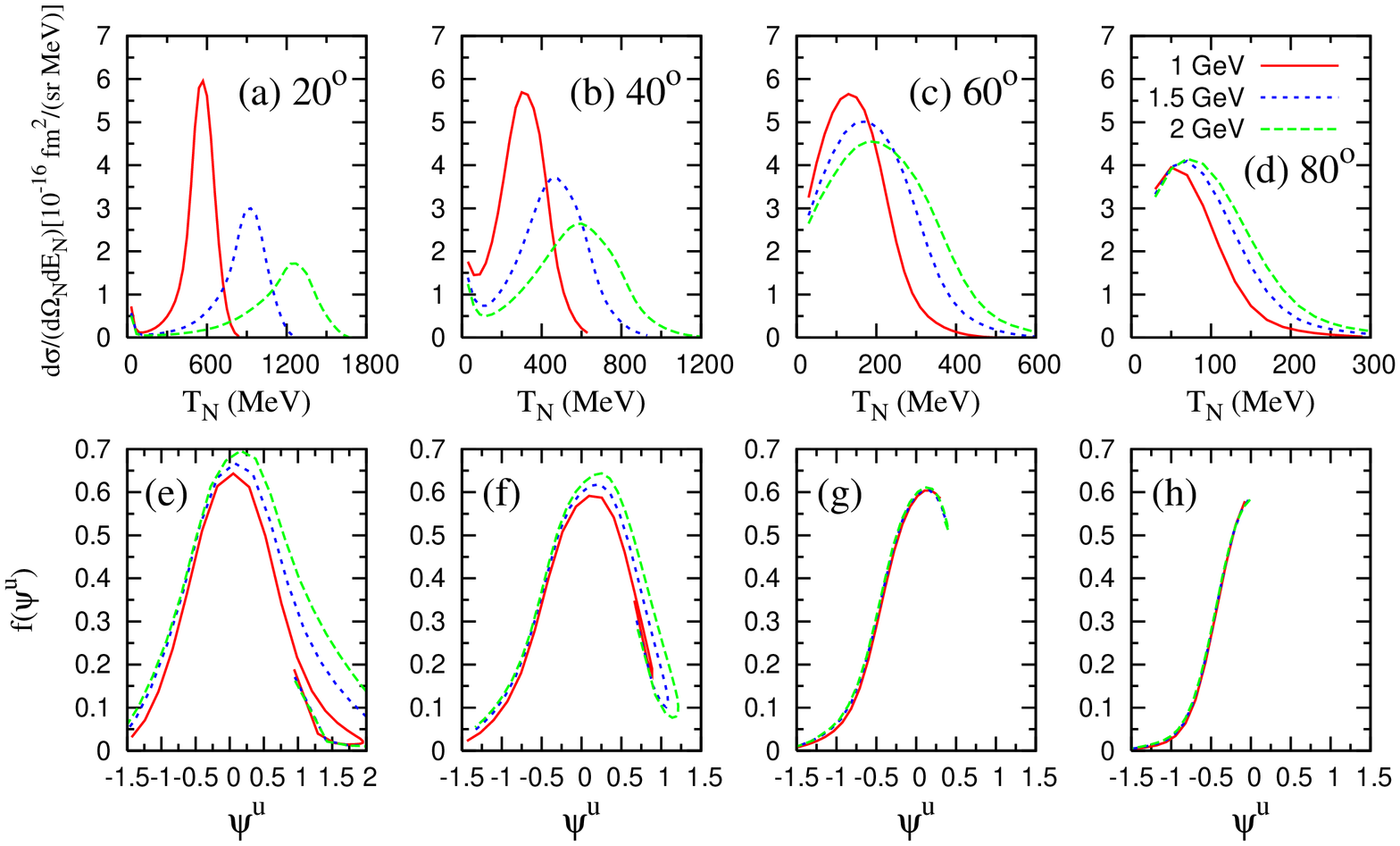}}} \par} \vspace{2cm}
\caption{(Color online) Differential cross section for $^{12}$C$(\nu,p)$ scattering in
RIA-RMF for different beam energies (panels (a), (b), (c) and (d)) and
their corresponding scaling functions (panels (e), (f), (g) and (h)). From left to right,
the graphs correspond to different values of $\theta_{kp_N}$, namely
20$^o$, 40$^o$, 60$^o$ and 80$^o$.}
\label{fig:firstkind_rmf}
\end{figure}

To begin let us focus on the first-kind scaling analysis of the
$(\nu,p)$ cross sections. In the top panels of
Fig.~\ref{fig:firstkind_rmf} we present $^{12}$C$(\nu,p)$ quasielastic 
differential cross sections for different nucleon detection angles
and beam energies. Each graph corresponds to a certain
$\theta_{kp_N}$, from 20$^o$ to 80$^o$, and for each
$\theta_{kp_N}$ value results for three
 neutrino beam energies, 1, 1.5, and 2 GeV, are displayed. The characteristic
momentum for $^{12}$C is taken to be $k_A=$ 228
MeV/$c$ as suggested by the superscaling analysis of electron scattering in~\cite{Maieron:2001it}. As already observed in
Fig.~\ref{fig:figura2} and explained from the simple consideration
of free nucleons in the previous section, the variation of the
cross sections is softer for angles equal to or larger than
60$^o$. For 40$^o$ and smaller there is a noticeable shift of the
cross sections to higher nucleon kinetic energies, as well as a
considerable variation in magnitude when the beam energy
increases. If first-kind scaling holds, all of these cross
sections should collapse to a unique curve when dividing by the neutral-current
single-nucleon cross section (see Eq.~(\ref{eq:supers_f})) and
plotting the results against the dimensionless scaling variable
$\psi^u$. These results are shown in the bottom panels of
Fig.~\ref{fig:firstkind_rmf}. Focusing on larger angles, namely
60$^o$ and 80$^o$, it is observed that scaling of first-kind is
fulfilled to a high degree even in the presence of the very strong
FSI included in the RIA-RMF model. In other words, the variations
in the cross sections observed for different neutrino energies are
well accounted for by the single-nucleon part of the cross
sections, which has been factored-out in obtaining the scaling
function. In contrast, for forward angles, 20$^o$ and 40$^o$,
first-kind scaling is clearly not so well respected. Nevertheless,
there are some features of first-kind scaling behaviour that still
persist for these forward angles: the peak of the scaling function
appears approximately at the same $\psi^u$ value for all beam
energies, and in the region of negative $\psi^u$-values first-kind
scaling is reasonably well respected. However, the scaling
functions obtained at 20$^o$ and 40$^o$
 show an increase in  the height of the peaks of the curves, as well as a shift to $\psi^u>0$
for increasing beam energy. This is similar to what is observed in RIA-RMF for
 the inclusive $(e,e')$ case. Actually, the experimental $(e,e')$ data do
leave room for some breaking of first-kind scaling in the region of
positive scaling variable.

\begin{figure}
{\par\centering
\resizebox*{0.6\textwidth}{0.27\textheight}{\rotatebox{270}
{\includegraphics{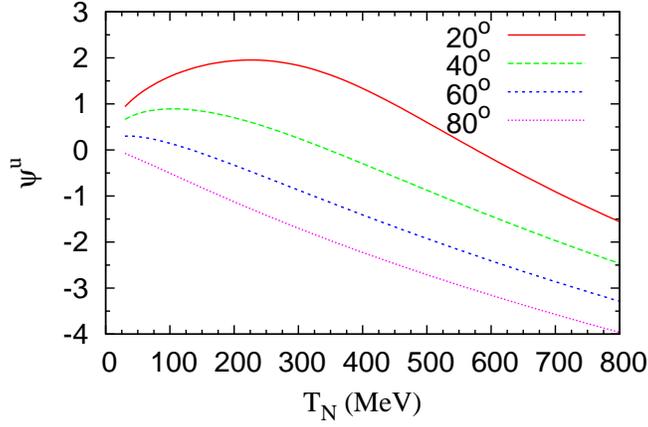}}} \par} \caption{(Color online) Behaviour of the
scaling variable $\psi^{u}$ as a function of the kinetic energy
of the outgoing nucleon at $\varepsilon=1$ GeV.} \label{fig:psi}
\end{figure}

A comment is in order concerning the behaviour of the scaling
function $f(\psi^u)$ for positive $\psi^u$ values. As noted, in
this $\psi^u$-region $f(\psi^u)$ may be two-valued. This purely
kinematical effect comes from the general expression for the
$\psi^u$ variable and the kinematics involved in neutral-current 
neutrino-nucleus scattering processes. This effect is present for
all models whenever the kinematics lead to positive
$\psi^u$-values close to or higher than 0.5 (most forward
scattering angles). In order to clarify this point we present in
Fig.~\ref{fig:psi} the behavior of the scaling variable $\psi^u$
at a beam energy of 1 GeV as a function of the outgoing nucleon
kinetic energy for four different values of the scattering angle,
$\theta_{kp_N}=20^o$, $40^o$, $60^o$ and $80^o$. As shown,
$\psi^u$ takes on a unique value for each $T_N$ for
$\theta_{kp_N}=60^o$ and $80^o$. For lower scattering angles,
there exists a region (wider for smaller $\theta_{kp_N}$) where
two different values of the nucleon kinetic energy lead to the
same $\psi^u$. Figure~\ref{fig:psi} also illustrates why the
superscaling functions in Fig.~\ref{fig:firstkind_rmf} extend up
to a different positive $\psi^u$-value depending on the nucleon
scattering angle. As a matter of fact, this $\psi^u$-value, that
corresponds to the minimum kinetic energy considered, is lower for
increasing angles.

\begin{figure}
{\par\centering
\resizebox*{1.1\textwidth}{0.55\textheight}{\rotatebox{270}
{\includegraphics{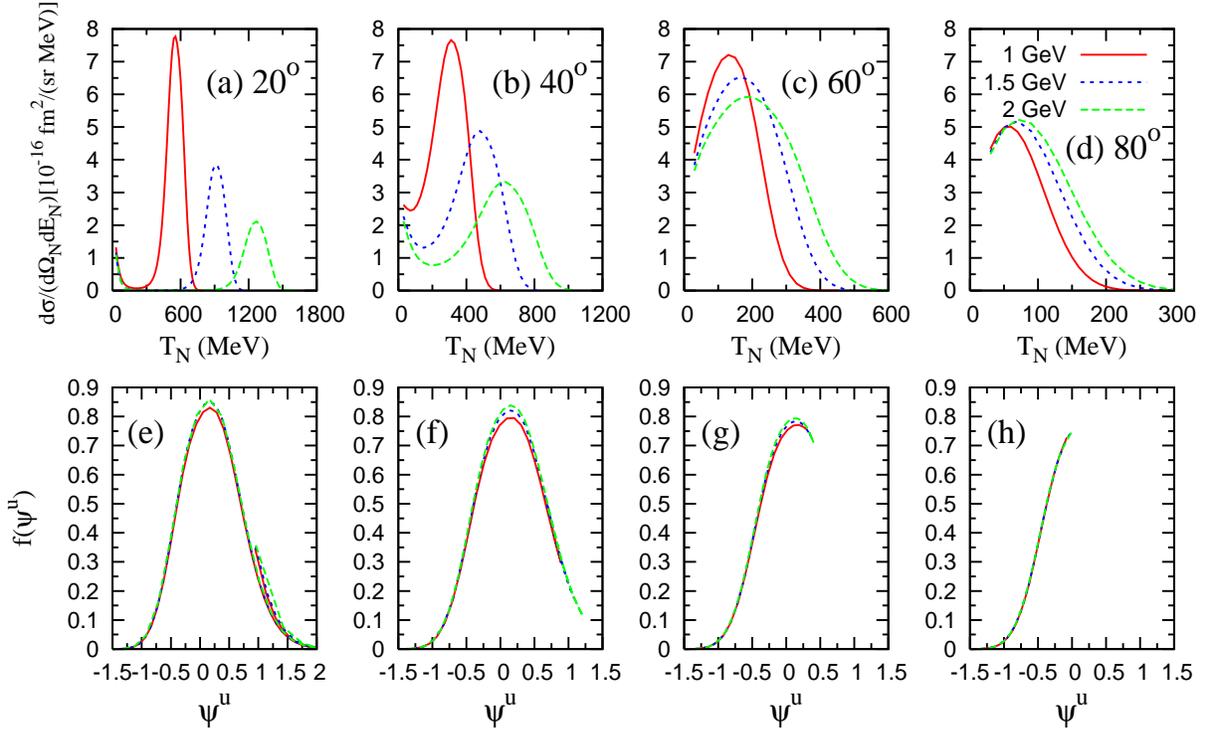}}} \par}
\vspace{2cm} \caption{(Color online) As in Fig.~\ref{fig:firstkind_rmf}, but now
for RPWIA.} \label{fig:firstkind_rpwia}
\end{figure}

\begin{figure}
{\par\centering
\resizebox*{0.75\textwidth}{0.75\textheight}{\rotatebox{0}
{\includegraphics{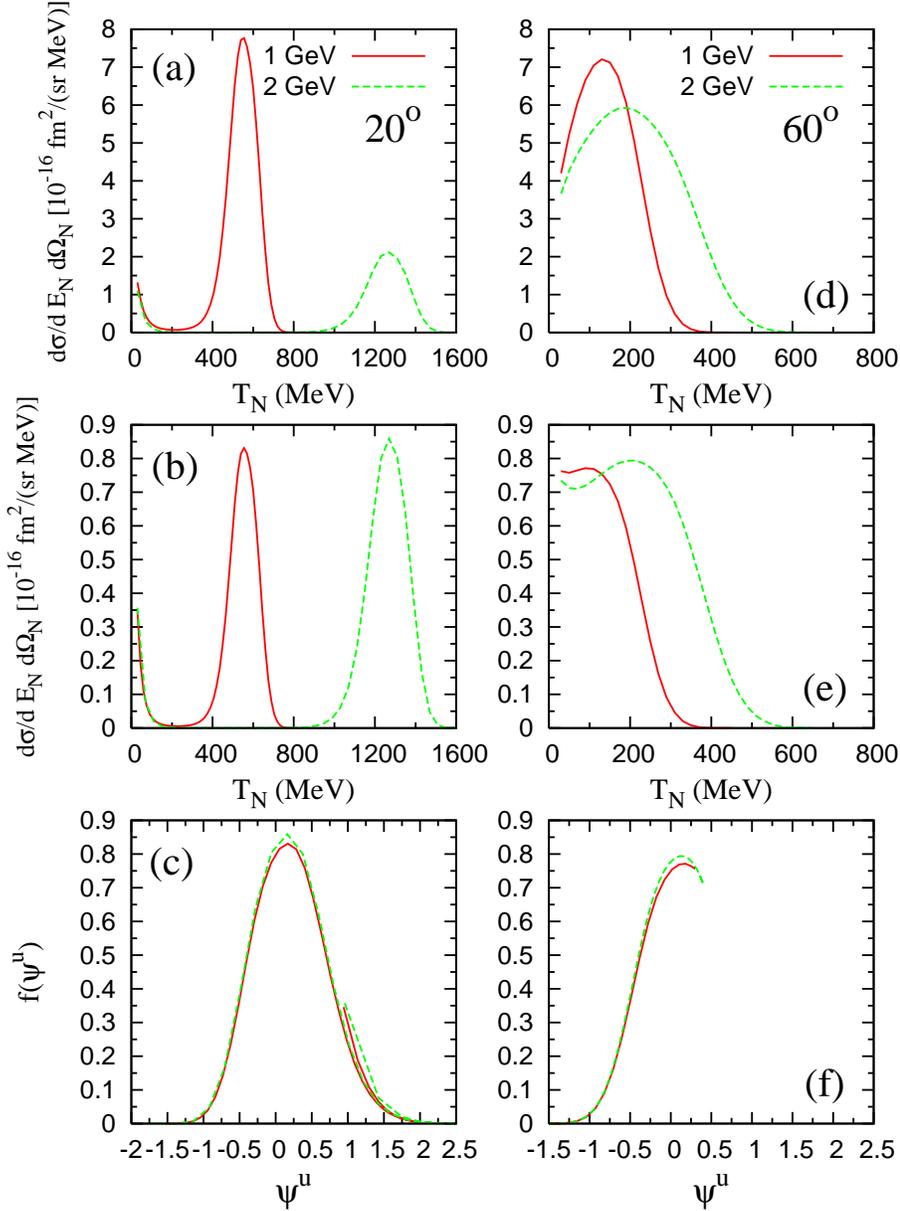}}} \par}
\caption{(Color online) Illustration of the scaling procedure for $\theta_{kp_N}=$20$^o$
 (panels (a), (b) and (c)) and $\theta_{kp_N}=$60$^o$ (panels (d), (e) and (f)) in the RPWIA.
 The top panels (a) and (d) show the differential cross sections against $T_N$ for
$^{12}$C$(\nu,p)$ at two beam energies for each angle. The middle
panels (b) and (e) show the
 results after factoring out the single-nucleon cross section and making
 dimensionless the results, namely the scaling function, but still as a function
 of $T_N$. The bottom panels (c) and (f) show the scaling function plotted against the scaling variable.}
\label{fig:scalingproc_rpwia}
\end{figure}

Up to now we have found that at forward scattering angles the
results display first-kind scaling violation to some extent, while
at larger angles exhibit almost perfect first-kind scaling. The
origin of this scaling behaviour of the QE neutral-current $(\nu,p)$ cross
sections is easily understood when one realizes that first-kind
scaling is very well fulfilled in the absence of final-state interactions,
 as was also
the case for electrons and charged-current neutrino
reactions~\cite{JA06,Caballero:2005sj,Alberico:1988bv,Amaro:2006pr}.
In Fig.~\ref{fig:firstkind_rpwia} we present results analogous to
those in Fig.~\ref{fig:firstkind_rmf}, but now for the RPWIA. In
contrast to what happens when FSI are accounted for, first-kind
scaling is now well fulfilled independently of the value of
$\theta_{kp_N}$. Therefore, the breakdown of first-kind scaling
for forward angles in Fig.~\ref{fig:firstkind_rmf} must be
ascribed to final-state interactions. This may appear to be a little unusual when one
examines the cross sections in the top panels in
Fig.~\ref{fig:firstkind_rpwia}, and makes a ``rough'' comparison
with the corresponding ones in Fig.~\ref{fig:firstkind_rmf}. The
variation with beam energy and nucleon scattering angle is very
similar in the two cases, and from this alone it is not clear why
the situations with and without FSI differ. In order to understand
clearly why and when FSI cause scaling violations, let us consider
the scaling procedure in two steps: i) dividing the cross sections
by the corresponding single-nucleon cross section (see
Eq.~(\ref{eq:supers_f})), and ii) representing the so-obtained
results against the scaling variable.
Figure~\ref{fig:scalingproc_rpwia} illustrates these two steps for
the cases of 20$^o$ (panels (a), (b) and (c)) and 60$^o$ (panels (d), (e) and (f)) in the RPWIA. Let us focus first on the top panels, where
we show the differential cross sections for $^{12}$C$(\nu,p)$ as a
function of the outgoing nucleon kinetic energy for two values of
neutrino beam energy. The differences between 20$^o$ and 60$^o$
cross sections are clearly visible. As already mentioned and
explained using the free-nucleon case in previous section, the
cross sections at 20$^o$ are subject to a much stronger shift to
higher $T_N$ values with increasing beam energy than at 60$^o$.
Also the magnitude of the cross sections at different beam
energies changes more drastically for 20$^o$ than for 60$^o$. If
we divide these differential cross sections by the corresponding
single-nucleon ones in order to get the scaling function, but
still plot the results against $T_N$, we obtain what is shown in
the middle panels of Fig.~\ref{fig:scalingproc_rpwia}. We observe
that this simple step already removes the differences in magnitude
for the different beam energies, not only for 60$^o$, but also for
20$^o$. If we now represent these results as a function of the
scaling variable, the shift observed when changing the beam energy
disappears, and scaling of the first kind shows up. Both steps, i)
and ii), must be carried out in order to bring together the cross
sections for different beam energies. We see how in RPWIA the
dependence of the neutral-current QE neutrino cross section on the energy of
the outgoing nucleon is well described by its single-nucleon
content, and it is consequently cancelled out during the scaling
analysis.

\begin{figure}
{\par\centering
\resizebox*{0.5\textwidth}{0.5\textheight}{\rotatebox{270}
{\includegraphics{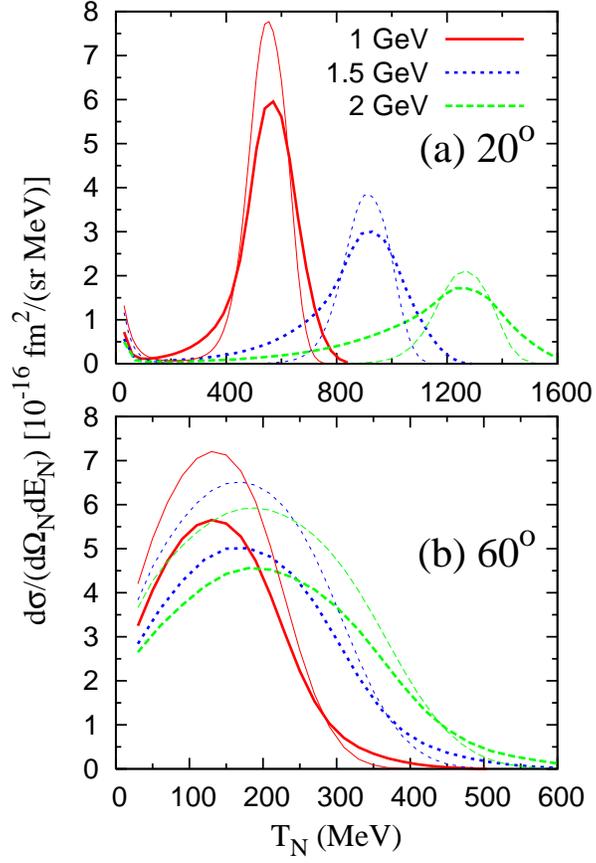}}} \par} \caption{(Color online) FSI effects in the differential
 cross sections for $^{12}$C$(\nu,p)$ scattering at three beam energies. The thick lines correspond to RMF calculations
 and the thin ones to RPWIA.
 Panel (a) refers to $\theta_{kp_N}=$20$^o$ and panel (b) to 60$^o$.} \label{fig:cross_rmfvsrpwia}
\end{figure}

\begin{figure}
{\par\centering
\resizebox*{0.75\textwidth}{0.75\textheight}{\rotatebox{0}
{\includegraphics{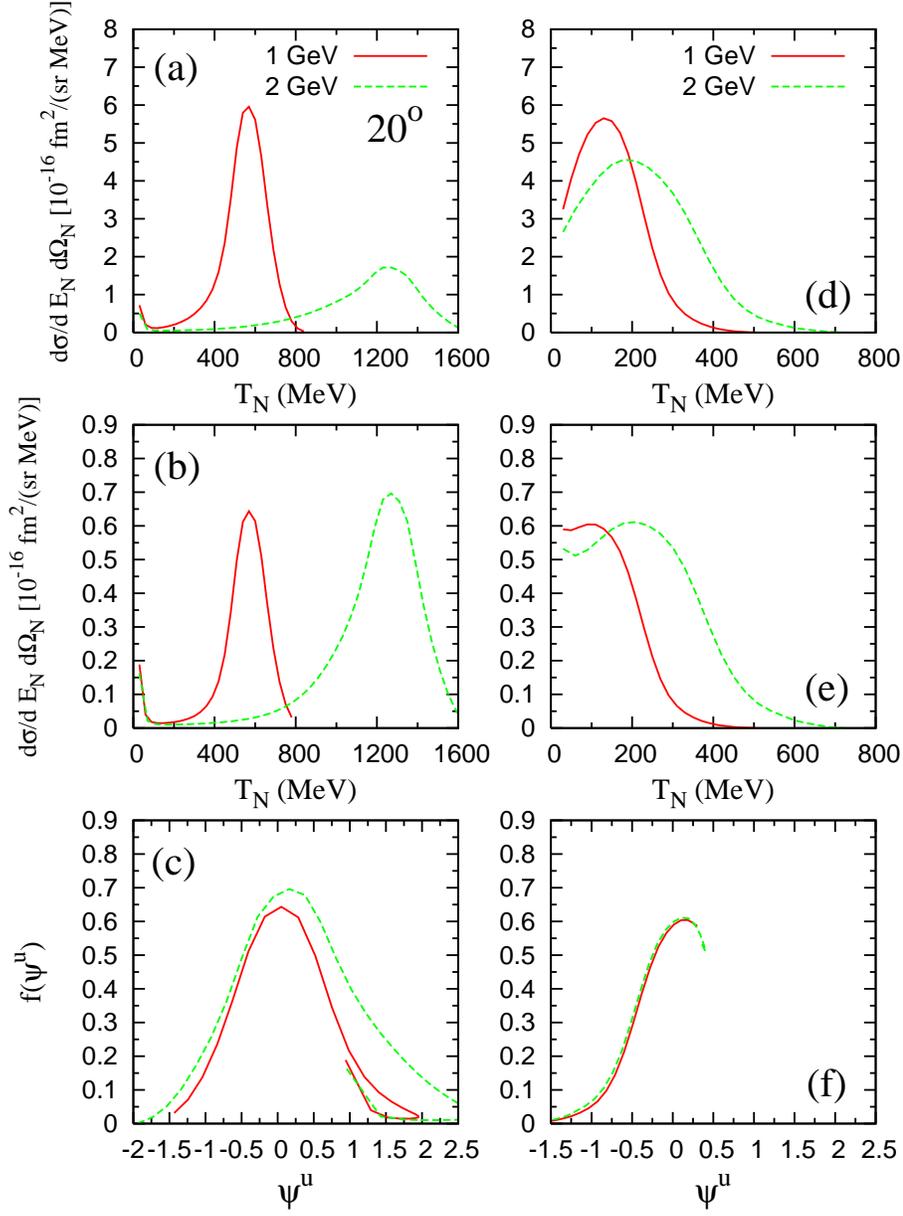}}} \par} \caption{(Color online) As in
Fig.~\ref{fig:scalingproc_rpwia}, but now for RMF.}
\label{fig:scalingproc_rmf}
\end{figure}

In what follows we analyze the situation when final-state interactions
 are included.
Figure~\ref{fig:cross_rmfvsrpwia} shows how FSI modify the
$^{12}$C$(\nu,p)$ cross sections at 20$^o$ and 60$^o$ for
different beam energies. These two angles are representative of
the regimes where first-kind scaling is broken (forward angles) or
fulfilled (large angles), respectively. One observes that FSI
involve a redistribution of strength that depends on the energy of
the final nucleon. Values of $T_N$ far from each other are
affected very differently by FSI. In the case of 20$^o$ this
effect is clearly stronger, as the cross sections for the various
beam energies span a broader region of nucleon kinetic energies.
This extra $T_N$-dependence due to final-state interactions is not contained in the
single-nucleon cross section, and thus the scaling procedure is
not enough to compensate for it if the cross sections peak at very
different $T_N$ values. This is clearly appreciated if we follow
steps i) and ii) as we have done previously in RPWIA. Results are
shown in Fig.~\ref{fig:scalingproc_rmf}. As in RPWIA, a large
amount of the variation in magnitude of the cross sections at
different beam energies is also removed when the single-nucleon
content is factored out, although clear differences still persist
in this case for 20$^o$. Moreover, at $20^o$ one observes the
additional $T_N$-dependence of the cross section with respect to
the plane-wave results (extended tails), and it consequently shows
up even when plotting the results against $\psi^u$. For $60^o$ the
behaviour of results differs because FSI effects are similar for
the different beam energies, given the fact that the cross
sections peak at similar $T_N$ values. In this case, we can say
that FSI effects do not spoil scaling. In general, if the
kinematics of the process are such that the range of energies of
the ejected nucleon depends strongly on the beam energy, the
nucleon may be subject to very different final-state interactions for each
$\varepsilon$, and a visible breakdown of first-kind scaling will
show up. This is what happens in general for forward angles, where
there is a strong shift of the position of the peak of the cross
section with incoming beam energy.
However, for those kinematics for which the range of $T_N$ remains
approximately the same when considering different beam energies,
{\it i.e,} for larger $\theta_{kp_N}$ (see
Fig.~\ref{fig:casolibre}), first-kind scaling emerges even with final-state
interactions included, as FSI effects on the knockout nucleon are similar for
different beam energies. This outcome, namely, almost perfect
first-kind scaling for large angles and violations in the case of
forward ones, is based on pure kinematical reasoning, and to this
extent (and within the impulse approximation) it is independent of the specifics of the
model employed. It appears for any model in which final-state interactions
 are strongly
$T_N$-dependent, as is the case of the relativistic mean-field description. One has to
realize that the use of the RMF potential provides a stringent
scenario for first-kind scaling in the case of QE neutral-current neutrino
reactions. If future QE neutral-current neutrino data suggest weaker final-state interactions, that
are less dependent on $T_N$ than the ones predicted by the relativistic-mean field, our
results imply that in such a case first-kind scaling would work
better for smaller angles than shown here, even without any
restriction on $\theta_{kp_N}$.

Concerning the bivalued nature of the scaling function
$f(\psi^u)$, Fig.~\ref{fig:firstkind_rpwia} shows that in the
absence of FSI superscaling is a good approximation and the two
values of the scaling function for the same $\psi^u$ value are
nearly equal. When FSI are present (Fig.~\ref{fig:firstkind_rmf}),
and, if the kinematics prevent superscaling, the bivalued
nature of the superscaled function is significantly revealed. 

\begin{figure}
{\par\centering
\resizebox*{0.6\textwidth}{0.27\textheight}{\rotatebox{270}
{\includegraphics{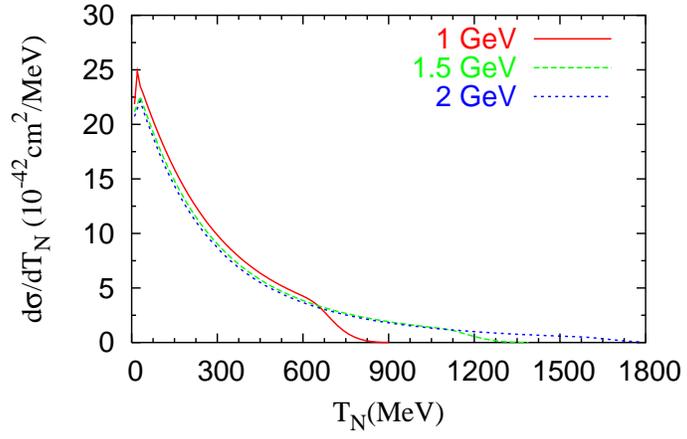}}} \par} \caption{(Color online) Differential cross section
integrated over angles for $^{12}$C$(\nu,p)$ scattering in
RIA-RMF for different beam energies.} \label{fig:dsigmadtn}
\end{figure}

It is very relevant that for the differential cross section
integrated over angles, namely $d\sigma/dT_N$, the larger
contributions come from angles for which scaling works nicely even
in the presence of very strong final-state interactions. This is clearly shown
from the comparison between $d\sigma/dT_N$ in
Fig.~\ref{fig:dsigmadtn} and the contributing
$d\sigma/dT_Nd\Omega_N$ cross sections in Fig.~\ref{fig:figura2}.
The situation is more favourable for higher beam energies, as in
this case the contribution coming from forward angles moves to
higher values of $T_N$, where the cross section has already
decreased considerably. Additionally, note that the contribution
of forward angles is suppressed in the integration by the
phase-space factor $\sin \theta_{kp_N}$.

\begin{figure}
{\par\centering
\resizebox*{1.1\textwidth}{0.55\textheight}{\rotatebox{270}
{\includegraphics{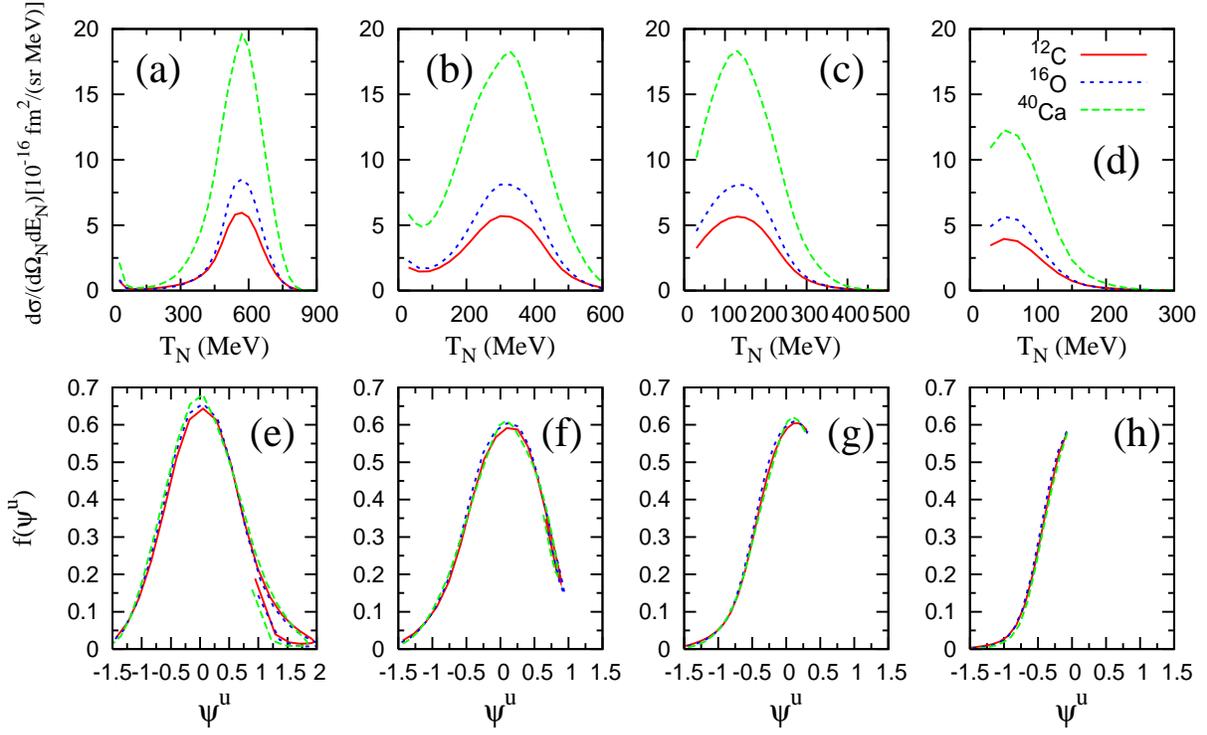}}} \par} \vspace{2cm}
\caption{(Color online) Differential cross section for $(\nu,p)$ scattering in
RIA-RMF at 1 GeV for different target nuclei (panels (a), (b), (c) and (d)) and
their corresponding scaling functions (panels (e), (f), (g) and (h)). As for
 Figs.~\ref{fig:firstkind_rmf} and~\ref{fig:firstkind_rpwia}, from left to
 right the graphs correspond to different values of $\theta_{kp_N}$, namely
20$^o$, 40$^o$, 60$^o$ and 80$^o$.}
\label{fig:scalingsecond}
\end{figure}

Results of a study of scaling of the second kind are presented in
Fig.~\ref{fig:scalingsecond}. The scaling function $f(q',\psi^u)$
is evaluated at $\varepsilon=1$ GeV for three different targets,
$^{12}$C as in previous figures, together with $^{16}$O and
$^{40}$Ca with $k_A=216$ and $241$ MeV/$c$, respectively. These values for
 the
characteristic momentum scale of $^{16}$O and
$^{40}$Ca provide the best superscaling of the data in the case of electron scattering~\cite{Maieron:2001it}. As in the
investigation of first-kind scaling, the values $\theta_{kp_N}=$
$20^o$, 40$^o$, 60$^o$ and 80$^o$ are considered. The superscaling
functions obtained for several nuclei are almost identical (bottom
panels), in spite of the strong difference in magnitude of the
corresponding cross sections (top panels). That is, the dependence
on the nuclear species is well accounted for by the superscaling
analysis. Scaling of second kind is seen to be very robust,
thereby opening up a means of taking into account nuclear effects
for different nuclei by employing superscaling ideas.

To sum up, superscaling for QE neutral-current neutrino reactions is fulfilled
at the level of $\sim$10$\%$ when very strong final-state interactions are accounted
for, provided the angle at which the nucleon is scattered is
larger than 50-60 degrees. This $\sim$10$\%$ criterion corresponds
to the maximum difference between the peaks of the scaling
functions when dealing with the different neutrino energies,
target nuclei and angles considered in this work.

\begin{figure}
{\par\centering
\resizebox*{1\textwidth}{0.55\textheight}{\rotatebox{270}
{\includegraphics{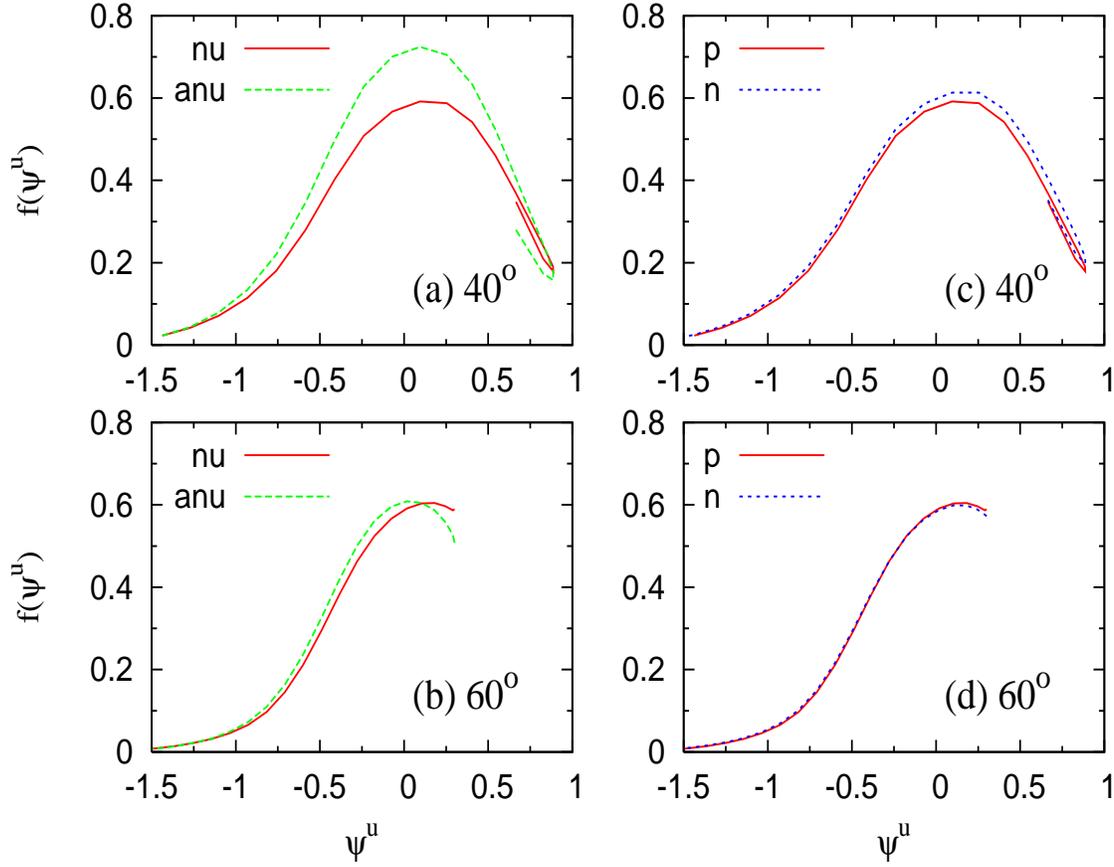}}} \par} \vspace{2cm}
\caption{(Color online) Scaling functions in RIA-RMF for $^{12}$C at 1 GeV for
$(\nu,p)$ versus $(\overline{\nu},p)$ in panels (a) and (b), and
for $(\nu,p)$ versus $(\nu,n)$ in panels (c) and (d). Panels (a) and (c) 
show results at $\theta_{kp_N}=$40$^o$, and (b) and (d)
correspond to
 $\theta_{kp_N}=$60$^o$.} \label{fig:nuvsanu}
\end{figure}

The above analysis has been focused on $(\nu,p)$ reactions.
Although not shown here in detail for the sake of brevity, a
similar study has been performed for $(\nu,n)$,
$(\overline{\nu},p)$ and $(\overline{\nu},n)$ cross sections,
reaching similar conclusions. In spite of the large differences of
magnitude and/or behaviour observed for cross sections for
neutrinos or antineutrinos, and for proton or neutron knockout,
the four processes converge to roughly the same superscaling
function, with some limitations that are commented upon next. In
particular, superscaling functions obtained for protons and
neutrons are almost identical in all cases considered in the
present work. This is illustrated in panels (c) and (d) of
Fig.~\ref{fig:nuvsanu} when considering scattering from $^{12}$C
at 1 GeV for two values of $\theta_{kp_N}$, 60 and 40 degrees,
that are representative of the kinematical regimes for which
first-kind scaling works well (large angles) or is moderately
violated (more forward ones). The agreement between superscaling
functions obtained for neutrinos and antineutrinos is somewhat
worse, as clearly shown in panels (a) and (b) of
Fig.~\ref{fig:nuvsanu}. In the case of $60^o$ there are
differences mainly in the region close to the peak of the scaling
functions, but overall the two curves are rather similar. However,
neutrino and antineutrino scaling functions show important
deviations for forward angles, for which the first-kind scaling is
not well fulfilled. Once more, the reason for this is clearly
linked to final-state interactions. We have checked that the $\nu$-$\overline{\nu}$
comparison improves considerably in the plane-wave limit.

The exhibition of superscaling by neutral-current quasielastic neutrino-nucleus
scattering even in the presence of very strong final-state interactions opens the door
for investigations of the validity of the universal character of
the scaling function for inclusive electroweak processes on
nuclei, using either electrons or charged- and neutral-current neutrino probes. To
the extent that this universality holds and consequently all of
these processes can be described by means of a unique scaling
function, the phenomenological SuSA approach formerly applied to
predict charged-current neutrino-nucleus cross sections could also provide
reliable, largely model-independent, predictions for neutral-current processes.

\begin{figure}
{\par\centering
\resizebox*{0.8\textwidth}{0.4\textheight}{\rotatebox{270}
{\includegraphics{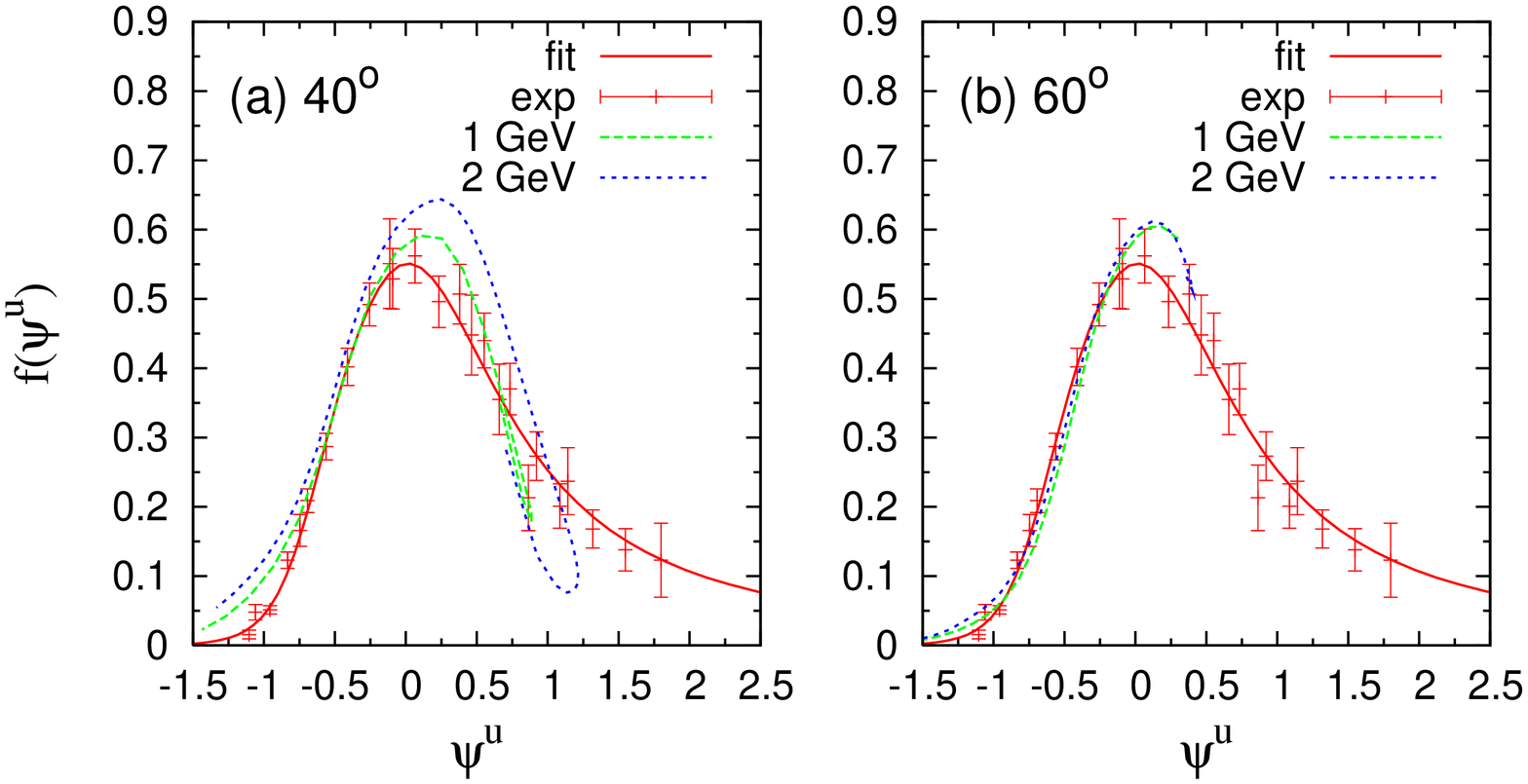}}} \par} \caption{(Color online) NC scaling function
 evaluated within the RIA-RMF approach for (a) 40 and (b) 60 degrees at two beam
 energies, compared with
 the averaged experimental function, together with a phenomenological
parameterization of the $(e,e')$ data.} \label{fig:expvsth}
\end{figure}

In order to study whether or not this universality assumption also
holds for neutral-current processes, in Fig.~\ref{fig:expvsth} we compare the
RIA-RMF NC superscaling function with the averaged experimental
function obtained from the analysis of quasielastic $(e,e')$ data, together
with a phenomenological
parameterization~\cite{Donnelly:1998xg,Maieron:2001it,Amaro:2004bs}. The
RIA-RMF superscaling function has been plotted for two values of
$\theta_{kp_N}$ for which scaling of first-kind is (a) well fulfilled
(60$^o$) or (b) not-so-well (40$^o$). Results are shown for two beam energies for each angle. As
observed, the model gives rise to a neutral-current scaling function that
follows closely the behaviour of the $(e,e')$ function whenever
superscaling is well respected (60$^o$). In such a case, it is
also noticed that the bivalued behaviour of the superscaling
function is hardly visible. In contrast, the departure from the
SuSA $(e,e')$ response is visible for the case of 40$^o$, for
which breakdown of first-kind scaling clearly occurs, and the
bivalued nature of the neutral-current superscaled function is enhanced. We
notice that all curves would coincide if superscaling were exactly
fulfilled in both neutral-current and $(e,e')$ cases. Since the $(e,e')$ and NC
scaling curves are obtained under rather different kinematical
situations, the scaling curves depart from one another when
superscaling is not a good approximation. This supports the
assumption that, under proper kinematical restrictions, a
universal quasielastic scaling function exists which is valid, not only for
inclusive electron and charged-current neutrino reactions as seen
in~\cite{JA06,Caballero:2005sj}, but also for neutral-current processes, making
feasible the idea of using the SuperScaling Approach also to predict neutral-current 
neutrino-nucleus cross sections.

\section{Conclusions}
\label{sec:conclusions}

In this paper we have studied the possibility of applying
superscaling ideas to neutral-current neutrino scattering processes, of
interest for experiments that rely on the measurement of
neutrino-nucleus cross sections. This is an extension of the work
presented in~\cite{cris07}. The shortage of neutral-current experimental data
to date has made mandatory the use of a specific model for this
purpose. We have chosen the RIA-RMF approach mainly because of its
capability to reproduce the superscaling properties exhibited by
inclusive electron data, including the asymmetric tail occurring
in the experimental scaling function that has proven to be elusive
for most models. Furthermore, RIA-RMF gives rise to scaling
violations that are similar to those presented by the electron
scattering data. The essential ingredient of the RIA-RMF model
that makes this agreement possible is the inclusion of strong final-state 
interactions 
between the emitted nucleon and the residual nucleus by means of
the same relativistic mean-field potential already used for
describing the initial bound nucleon states. We have only relied
on the RIA-RMF model to illustrate possible FSI effects in the
superscaling properties of quasielastic neutral-current neutrino cross sections. Our goal
was not to explore how this particular model (super-) scales or
not, but rather whether and when the experimental scaling response
obtained from electron scattering data can be applied to predict
neutral-current neutrino-nucleus cross sections, for which we consider the
RIA-RMF model as guidance.

Within the context of the RIA-RMF, we have evaluated inclusive
differential cross sections (and separate response functions) for
various choices of kinematics and nuclei. The superscaling
function has been computed and displayed as a function of the
scaling variable $\psi^u$ for the different kinematics and various
target nuclei. Proceeding in this way, we have investigated
scaling of first and second kinds. From our RIA-RMF results,
scaling of the second kind is seen to work extremely well, opening
the possibility of accounting for nuclear effects for different
nuclei employing superscaling ideas. With respect to scaling of
first kind, there are kinematics for which scaling is very good
and others for which clear scaling violations are observed. We
have identified what are the conditions that make a specific
choice of kinematics suitable or not for good first-kind scaling:
the kinematics  must be such that the range of energies spanned by
the ejected nucleon depends weakly on the incoming neutrino
energy. If the beam energy is given and the angle of the ejected
nucleon with respect to the beam is fixed, the above situation
means angles larger than roughly 50 degrees, which happens to be
the region where the cross section integrated over angles reaches
larger values. In such cases, first-kind scaling is well respected
even in the presence of strong final-state interactions.

We have compared the RIA-RMF superscaling function with the
experimental $(e,e')$ scaling function and have verified that the
agreement is good for those kinematics for which first-kind
scaling is well fulfilled. This accordance clearly favours the
existence of a universal scaling function which is valid for
electron and neutrino (charged- or neutral-current) inclusive scattering, and it
gives us confidence that, under the kinematical restrictions
explained above that ensure good first-kind scaling, the SupersScaling Approach
 can be
extended to predict neutral-current quasielastic neutrino cross sections. We also note
that, even though we have illustrated this study within the
RIA-RMF model, the kinematical conditions that grant the validity
of SuSA are model independent provided the impulse approximation can 
be safely
applied, that is, under quasielastic kinematics with neutrino beam energies
from $\sim$500 MeV up to a few GeV. Obviously, there also exist
some uncertainties in the assumption of the validity of the
(super)scaling hypothesis for neutral currents, mostly due to the
fact that meson exchange currents have not been
considered. Meson-exchange current effects do not scale (see~\cite{mec1,mec2}), but
they are expected to contribute less than 15$\%$ (see for
instance~\cite{mec3,mec4}) to these inclusive neutrino-nucleus
cross sections at intermediate energies.


\section*{Acknowledgements}
This work was partially supported by DGI (Spain) and FEDER funds:
 FIS2005-01105, FPA2006-07393, FPA2006-13807 and FPA2007-62216, by the Junta de
 Andaluc\'{\i}a and the INFN-CICYT, and by the Comunidad de Madrid and UCM
 (910059 `Grupo de F\'{\i}sica Nuclear' and PR1/07-14895). It was
also supported in part (TWD) by U.S. Department of Energy Office of
Nuclear Physics under contract No. DE-FG02-94ER40818. M.C.M
acknowledges a `Juan de la Cierva' contract from MEC. Computations
were performed at the `High Performance Cluster for Physics'
  funded by UCM and FEDER funds.



\begin{thebibliography}{99}

\bibitem{neut_exp}
Y. Fukuda {\it et al.} (Super-Kamiokande Col.), Phys. Rev.
     Lett. {\bf 81}, 1562 (1998), M. H. Ahn {\it et al.} (K2K),
     Phys. Rev. Lett. {\bf 90}, 041801 (2003), E. Ables {\it et al.} (MINOS),
     Fermilab-Proposal-0875 (1995), A. A. Aguilar-Arevalo
     {\it et al.} (MiniBooNe), Phys. Rev. Lett. {\bf 98}, 231801 (2007),
     I. Ambats {\it et al.} (NOvA.),
     Fermilab-Proposal-0929 (2004), D. Drakoulakos {\it et al.} (MINERvA), hep-ex/0405002, S. Brice {\it et al.} (FINeSSE), hep-ex/0402007.

\bibitem{Amaro:2004bs}
  J.~E.~Amaro, M.~B.~Barbaro, J.~A.~Caballero, T.~W.~Donnelly, A.~Molinari and I.~Sick,
  Phys.\ Rev.\ C {\bf 71}, 015501 (2005).

\bibitem{Caballero:2005sj}
  J.~A.~Caballero, J.~E.~Amaro, M.~B.~Barbaro, T.~W.~Donnelly, C.~Maieron and J.~M.~Ud\'{\i}as,
  Phys.\ Rev.\ Lett. {\bf 95}, 252502 (2005).

\bibitem{Amaro:2005dn}
  J.~E.~Amaro, M.~B.~Barbaro, J.~A.~Caballero, T.~W.~Donnelly and C.~Maieron,
  Phys.\ Rev.\  C {\bf 71}, 065501 (2005).

\bibitem{JA06}
J.~A.~Caballero,
Phys. Rev. C {\bf 74}, 015502 (2006).

\bibitem{Amaro:2006if}
  J.~E.~Amaro, M.~B.~Barbaro, J.~A.~Caballero, T.~W.~Donnelly and J.~M.~Ud\'{\i}as,
  Phys.\ Rev.\  C {\bf 75}, 034613 (2007).


\bibitem{Antonov:2006md}
  A.~N.~Antonov {\it et al.},
  Phys.\ Rev.\  C {\bf 74}, 054603 (2006).

\bibitem{Martini:2007jw}
  M.~Martini, G.~Co', M.~Anguiano and A.~M.~Lallena,
  Phys.\ Rev.\  C {\bf 75}, 034604 (2007).

\bibitem{Amaro:2006tf}
  J.~E.~Amaro, M.~B.~Barbaro, J.~A.~Caballero and T.~W.~Donnelly,
  Phys.\ Rev.\ Lett.\  {\bf 98}, 242501 (2007).

\bibitem{Caballero:2007tz}
  J.~A.~Caballero, J.~E.~Amaro, M.~B.~Barbaro, T.~W.~Donnelly and J.~M.~Ud\'{\i}as,
Phys.\ Lett.\ B {\bf 653}, 366 (2007).

\bibitem{Amaro:2006pr}
  J.~E.~Amaro, M.~B.~Barbaro, J.~A.~Caballero and T.~W.~Donnelly,
  Phys.\ Rev.\  C {\bf 73}, 035503 (2006)

\bibitem{Antonov2007_1}
A.~N.~Antonov, M.~V.~Ivanov, M.~B.~Barbaro, J.~A.~Caballero, E.~M.~de Guerra, and M.~K.~Gaidarov,
Phys.\ Rev.\ C {\bf 75}, 064617 (2007).

\bibitem{Day:1990mf}
  D.~B.~Day, J.~S.~McCarthy, T.~W.~Donnelly and I.~Sick,
  Ann.\ Rev.\ Nucl.\ Part.\ Sci.\  {\bf 40}, 357 (1990).

\bibitem{Donnelly:1998xg}
  T.~W.~Donnelly and I.~Sick,
  Phys.\ Rev.\ Lett.\  {\bf 82}, 3212 (1999).

\bibitem{Donnelly:1999sw}
  T.~W.~Donnelly and I.~Sick,
  Phys.\ Rev.\ C {\bf 60}, 065502 (1999).

\bibitem{Maieron:2001it}
C.~Maieron, T.~W.~Donnelly and I.~Sick, Phys.\ Rev.\ C {\bf 65},
025502 (2002).

\bibitem{Barbaro:1998gu}
  M.~B.~Barbaro, R.~Cenni, A.~De Pace, T.~W.~Donnelly and A.~Molinari,
  Nucl.\ Phys.\ A {\bf 643}, 137 (1998).


\bibitem{Barbaro:2003ie}
M.~B.~Barbaro, J.~A.~Caballero, T.~W.~Donnelly and C.~Maieron,
Phys.\ Rev.\ C {\bf 69},  035502 (2004).

\bibitem{peterson}
  R.~J.~Peterson,
  Nucl.\ Phys.\  A {\bf 769}, 95 (2006); {\bf 769}, 115 (2006); {\bf 791}, 84 (2007); {\bf 803}, 46 (2008)

\bibitem{Alberico:1988bv}
  W.~M.~Alberico, A.~Molinari, T.~W.~Donnelly, E.~L.~Kronenberg and J.~W.~Van Orden,
  Phys.\ Rev.\ C {\bf 38}, 1801 (1988).

\bibitem{Barbaro:1996vd}
  M.~B.~Barbaro, A.~De Pace, T.~W.~Donnelly, A.~Molinari and M.~J.~Musolf,
  Phys.\ Rev.\ C {\bf 54}, 1954 (1996).

\bibitem{Udias}
J.M.~Ud\'{\i}as {\it et al.},
Phys.\ Rev.\ C {\bf 48}, 2731 (1993);
{\bf 51}, 3246 (1995);
{\bf 64}, 024614-1 (2001).

\bibitem{Alb97}
W.~M.~Alberico {\it et al.},
Nucl.\ Phys.\ A {\bf 623}, 471 (1997);
Phys.\ Lett.\ B {\bf 438}, 9 (1998);
Nucl.\ Phys.\ A {\bf 651}, 277 (1999).

\bibitem{Martinez:2005xe}
  M.~C.~Mart\'{\i}nez, P.~Lava, N.~Jachowicz, J.~Ryckebusch, K.~Vantournhout and J.~M.~Ud\'{\i}as,
  Phys.\ Rev.\  C {\bf 73}, 024607 (2006)

\bibitem{boundwf}
C.J.~Horowitz and B.D.~Serot,
Nucl.\ Phys.\ A {\bf 368}, 503 (1981);
Phys.\ Lett.\ B {\bf 86}, 146 (1979);
%
B.D.~Serot and J.D.~Walecka, Adv.\ Nucl.\ Phys.\ {\bf 16}, 1 (1986).
%

\bibitem{Caballero:1997gc}
  J.~A.~Caballero, T.~W.~Donnelly, E.~Moya de Guerra and J.~M.~Ud\'{\i}as,
  Nucl.\ Phys.\  A {\bf 632}, 323 (1998).

\bibitem{Caballero:1998ip}
  J.~A.~Caballero, T.~W.~Donnelly, E.~Moya de Guerra and J.~M.~Ud\'{\i}as,
  Nucl.\ Phys.\  A {\bf 643}, 189 (1998)


\bibitem{Chiara03}
C. Maieron, M.C. Mart\'{\i}nez, J.A. Caballero and J.M. Ud\'{\i}as,
Phys. Rev. C {\bf 68}, 048501 (2003).

\bibitem{cris07}
  M.~C.~Mart\'{\i}nez, J.A.~Caballero, T.W.~Donnelly and J.~M.~Ud\'{\i}as,
 Phys.\ Rev.\ Lett.\  {\bf 100}, 052502 (2008).
%
\bibitem{Jin92}
Y.~Jin, D.~S.~Onley, and L.~E.~Wright,
  Phys.\ Rev.\  C {\bf 45}, 1333 (1992).
%
\bibitem{plohl}
O.~Plohl, C.~Fuchs, and E.N.E.~van Dalen,
 Phys. Rev. C {\bf 73}, 014003 (2006).
%
\bibitem{hori}
Y.~Horikawa, F.~Lenz, and Nimai C. Mukhopadhyay,
 Phys. Rev. C {\bf 22}, 1680 (1980).
%
\bibitem{chinn}
C.R.~Chinn, A.~Picklesimer, and J.W.~Van Orden,
 Phys. Rev. C {\bf 40}, 790 (1989).
%
\bibitem{meucci03}
A.~Meucci, F.~Capuzzi, C.~Giusti, and F.D.~Pacati,
 Phys. Rev. C {\bf 67}, 054601 (2003).
%
\bibitem{meucci04}
A.~Meucci, C.~Giusti, and F.D.~Pacati,
Nucl.\ Phys.\  A {\bf 739}, 277 (2004)
%
\bibitem{mec1}
  J.~E.~Amaro, M.~B.~Barbaro, J.~A.~Caballero, T.~W.~Donnelly and A.~Molinari,
  Phys.\ Rept.\  {\bf 368}, 317 (2002).
%
\bibitem{mec2}
  J.~E.~Amaro, M.~B.~Barbaro, J.~A.~Caballero, T.~W.~Donnelly and A.~Molinari,
  Nucl.\ Phys.\  A {\bf 723}, 181 (2003).
%
\bibitem{mec3}
  Y.~Umino, J.~M.~Ud\'{\i}as, and P.J.~Mulders,
 Phys.\ Rev.\ Lett.\  {\bf 74}, 4993 (1995).
%
\bibitem{mec4}
  Y.~Umino and J.~M.~Ud\'{\i}as,
 Phys.\ Rev.\  C {\bf 52}, 3399 (1995).
%
\end{thebibliography}
\end{document}